\documentclass[journal]{IEEEtran} 

\usepackage{amsmath, amsfonts, amssymb, mathtools}
\usepackage{graphicx}
\usepackage{tabularx}
\usepackage{array}
\usepackage{textcomp}
\usepackage{stfloats}
\usepackage{url}
\usepackage{booktabs}
\usepackage{verbatim}
\usepackage{float}
\usepackage{color}
\usepackage{cite}
\usepackage{bm}
\usepackage{bbm}   
\usepackage{tikz}
\usetikzlibrary{arrows.meta,positioning,fit,shapes.geometric}
\usepackage{algorithm}
\usepackage{algpseudocode}

\renewcommand{\arraystretch}{1.05}
\usepackage{xcolor}
\definecolor{myblue}{RGB}{0,90,180}
\definecolor{mygreen}{RGB}{0,130,0}
\colorlet{BLUE}{blue}

\usepackage[font=footnotesize]{caption}
\usepackage{subcaption}
\usepackage{hyperref}
\hypersetup{
	colorlinks=true,
	citecolor=blue,
	linkcolor=blue,
	urlcolor=blue,
	pdftitle={Secret Communications for UAVs with Movable Antennas in SAGIN},
	pdfauthor={Jiayang Wan, Yafei Wang, Jiawei Zhuang, Wenjin Wang, and Shi Jin},
	pdfsubject={CKM-enabled multi-agent reinforcement learning for secrecy energy efficiency}
}

\makeatletter
\renewcommand*{\eqref}[1]{%
	\hyperref[{#1}]{\textup{\tagform@{\ref*{#1}}}}%
}
\makeatother
\begin{document}
\bstctlcite{IEEEexample:BSTcontrol}
	
\title{Toward Secure Communications for a UAV Swarm with Movable Antennas in SAGIN:  CKM-Enabled Multi-Agent Reinforcement Learning Framework}

	\author{ 
   	Jiayang Wan, \IEEEmembership{Graduate Student Member, IEEE}, 
Yafei Wang, \IEEEmembership{Graduate Student Member, IEEE},
\\
Jiawei Zhuang, \IEEEmembership{Graduate Student Member, IEEE},
        Wenjin Wang, \IEEEmembership{Member, IEEE},
		  Tony Q.S. Quek, \IEEEmembership{Fellow, IEEE}

\thanks{
		 Jiayang Wan, Yafei Wang, Jiawei Zhuang and Wenjin Wang are with the National Mobile Communications Research
		Laboratory, Southeast University, Nanjing 210096, China, and also
		with Purple Mountain Laboratories, Nanjing 211100, China (e-mail:
        \{jywan, wangyf, jwzhuang, wangwj\}@seu.edu.cn).} 
      \thanks{
      	Tony Q. S. Quek is with the Information Systems Technology and Design Pillar, Singapore University of Technology and Design, Singapore 487372, and also with the Yonsei Frontier Laboratory, Yonsei University, Seoul 03722, South Korea (e-mail: tonyquek@sutd.edu.sg).}  
	}

\markboth{}{}
\pagestyle{headings}

\maketitle
\thispagestyle{headings}

\begin{abstract}

Space-air-ground integrated networks (SAGINs) can provide ubiquitous and reliable connectivity for unmanned aerial vehicles (UAVs). However, air-to-ground links, which are typically dominated by line-of-sight (LoS) propagation, are vulnerable to passive eavesdropping due to the broadcast nature of wireless channels. 
To enhance physical-layer security, we investigate a SAGIN-enabled secure downlink communication system in which UAVs select service links among satellite, aerial, and terrestrial networks while adjusting the positions of the movable antenna (MA) array to fully exploit connectivity and spatial degrees of freedom for improved secrecy communication performance.
Specifically, we maximize the secrecy energy efficiency (SEE) of a UAV swarm by jointly optimizing the MA positions, UAV trajectories, and link selections, subject to UAV mobility, MA movement, and link connectivity constraints.
To reduce the real-time channel state information (CSI) acquisition overhead, we propose a channel knowledge map (CKM)-assisted multi-agent reinforcement learning  framework. Specifically, the CKM is first constructed from sparse channel measurements via Kriging interpolation and is then leveraged together with satellite ephemeris information to enable efficient storage and retrieval of CSI.
To reduce the action-space dimensionality and computational complexity, we model the MA array using rigid-body kinematics and adjust its position through global rigid-body translation, thereby constructing a low-dimensional hybrid action space for the joint optimization decisions.
To align local decisions with system-wide performance under system constraints, we design an individual-team collaborative reward mechanism and introduce action masks to enforce constraints on UAV mobility, collision avoidance,  MA regions, and connectivity capacity.
Simulation results demonstrate that the proposed framework effectively improves the secrecy energy efficiency of the UAV swarm without incurring additional online CSI signaling overhead.

	\end{abstract}
	

	\vspace{-2mm}
\section{Introduction}\label{Introduction}
	\vspace{-1mm}

\IEEEPARstart{T}{h}e low-altitude economy, as an emerging and integrated economic paradigm, encompasses a wide spectrum of low-altitude flight activities and associated industrial applications conducted in airspace below 1000 meters \cite{wang2024sustainable}. 
Existing terrestrial networks (TNs)  cannot ensure continuous and reliable coverage in low-altitude airspace characterized by significant altitude variations and high horizontal mobility, resulting in pronounced coverage gaps \cite{wang2025unified}\cite{wan2025qos}.
With the evolution toward sixth-generation (6G) wireless networks, space-air-ground integrated networks (SAGINs), enabled by the deep integration of TNs, low-altitude wireless networks (LAWNs), and satellite networks (SNs), are emerging as a key technology for supporting future ubiquitous connectivity, with the potential to provide truly pervasive communication coverage for low-altitude airspace \cite{wang2025mobile}\cite{qiao2022joint}\cite{wang2026statistical}.

However, the high mobility of unmanned aerial vehicles (UAVs) leads to rapidly varying link geometries and propagation conditions, posing significant challenges to accurate and timely link selection in heterogeneous networks~\cite{amer2020mobility}.
Although trajectory optimization can partially mitigate these effects, it operates at the macroscopic position level and thus cannot fully exploit local spatial channel variations. Moreover, conventional fixed-position antenna (FPA) arrays, constrained by their fixed geometries, have limited ability to exploit additional spatial diversity~\cite{zhu2023modeling}.
Movable antenna (MA) technology addresses this limitation by repositioning antenna elements within confined regions to exploit additional spatial degrees of freedom and attain more favorable channel conditions \cite{zhu2023modeling}\cite{cao2025channel}. Related reconfigurable antenna technologies include fluid antennas \cite{zhu2024historical} and flexible antennas \cite{zheng2024flexible}.
On UAV platforms, MA arrays add fine-grained spatial reconfigurability to macro-scale UAV mobility, enabling adaptive reshaping of spatial channel responses and performance gains over conventional FPA arrays \cite{shao2025network}.
Building on this, recent studies have increasingly explored the use of UAV swarms for security patrols, military reconnaissance, and confidential data transmission~\cite{lei2025multi}. However, air-to-ground links are typically dominated by line-of-sight (LoS) propagation, and the broadcast nature of wireless channels makes them vulnerable to passive eavesdropping~\cite{jiang2021covert}. Therefore, cooperative secure communication for UAV swarms is of significant importance.

 \vspace{-3mm}

\subsection{Related Work}\label{Related Works}

Existing studies primarily improve UAV service continuity in heterogeneous networks through link selection and trajectory optimization. Depending on whether link-selection decisions are made within a homogeneous network or across heterogeneous networks, link selection can be classified into intra-network \cite{zhang2024toward}\cite{yang2023dqn} and inter-network selection \cite{xu2017modeling}\cite{liu2023user}\cite{wang2024sustainable}. 
 Intra-network link selection focuses on association and handover among nodes of the same network type. In TNs, handover parameters, including hysteresis, time-to-trigger (TTT), and cell individual offset (CIO), are optimized, while learning-based schemes are employed to reduce frequent and redundant handovers~\cite{zhang2024toward}. In SNs, predictable orbital and ephemeris information is exploited to optimize satellite association and reduce handover-related signaling overhead~\cite{yang2023dqn}. 
 By contrast, inter-network link selection coordinates TNs, LAWNs, and SNs with distinct propagation characteristics, coverage ranges, and link dynamics. Existing studies \cite{xu2017modeling}\cite{liu2023user}\cite{wang2024sustainable} have adopted stochastic-geometry modeling \cite{xu2017modeling}, TN-NTN coordination \cite{liu2023user}, and graph-based mobility management \cite{wang2024sustainable} to improve service continuity and avoid unnecessary handovers.
Since the UAV position directly affects the propagation conditions of candidate links, trajectory optimization and link selection are inherently coupled. Conventional studies \cite{zhang2018cellular}\cite{bulut2018trajectory}\cite{zhang2019trajectory} optimize UAV trajectories under connectivity constraints using graph-theoretic methods \cite{zhang2018cellular}, convex optimization \cite{bulut2018trajectory}, or dynamic programming \cite{zhang2019trajectory} to reduce flight time and outage probability or limit the maximum disconnection duration. 
To overcome the reliance of conventional optimization methods on accurate system models, recent studies \cite{zeng2021simultaneous}\cite{zhan2022energy} have employed reinforcement learning (RL) to optimize UAV trajectories and mission execution, thereby improving long-term connectivity reliability and energy efficiency.
However, existing approaches rely primarily on macro-scale UAV motion and fixed-position antenna arrays, leaving the spatial channel reconfigurability of MAs unexplored.

To overcome the limited ability of FPA arrays to exploit local spatial channel variations, MAs and FASs reconfigure antenna positions within confined regions to obtain more favorable channel conditions and avoid deep fading~\cite{zhu2024movable}. Six-dimensional movable antennas (6DMAs) further introduce rotational degrees of freedom, thereby enhancing spatial channel reconfigurability~\cite{shao2025network}.
Existing studies~\cite{zhu2023movable}\cite{bai2025movable}\cite{liu2025uav_dup1} have demonstrated the potential of MAs to improve array gain and multiple-input multiple-output (MIMO) capacity, and have begun to explore their integration with UAV platforms. For example, \cite{bai2025movable} optimizes the positions of directional MAs mounted on a UAV to reduce the data-collection time in backscatter sensor networks, whereas \cite{liu2025uav_dup1} deploy MA arrays on UAVs to improve the minimum array gain and system sum rate, respectively.
Beyond improving transmission performance, integrating MAs with UAVs can also enhance communication security. Existing studies~\cite{wen2026flexible}\cite{kim2026energy} jointly optimize UAV trajectories, transmission strategies, and MA positions to improve secrecy communication performance. However, these studies are largely limited to single-UAV or single-network scenarios~\cite{kim2026energy}, without jointly considering link selection, UAV trajectories, and MA configurations for multi-UAV SAGINs. Moreover, their reliance on instantaneous channel state information (CSI) incurs considerable signaling and computational overhead.

To reduce the overhead of real-time CSI acquisition, channel knowledge maps (CKMs) integrate offline channel measurements, location information, and environmental features to enable low-overhead online channel querying and prediction~\cite{wan2026channel}\cite{zeng2021toward}.
Meanwhile, multi-agent reinforcement learning (MARL) leverages inter-agent cooperation to address long-horizon sequential decision-making in multi-UAV systems and has been applied to cooperative trajectory design, task scheduling, and communication mode selection~\cite{hu2020cooperative}\cite{wu2020cellular}.
However, existing CKM and MARL studies have largely evolved independently: the former primarily supports channel prediction, whereas the latter typically relies on online CSI for cooperative task execution in UAV swarms.

\vspace{-3mm}

\subsection{Motivation and Main Contributions}\label{Motivation and Main Contributions}

Although significant progress has been made in link selection, UAV trajectory optimization, and MA-enabled communications, their joint optimization for secure UAV communications in SAGINs remains largely unexplored. Specifically, MA position optimization, UAV trajectory optimization, and link selection all rely on timely and accurate CSI, whose acquisition in SAGINs incurs substantial signaling overhead. Moreover, the strong coupling among these optimization variables, together with constraints on UAV mobility, MA movement ranges, and link connectivity, significantly increases the complexity of the joint optimization problem and may hinder algorithmic convergence. 
Meanwhile, cooperative control of a UAV swarm inherently introduces a high-dimensional trajectory decision space, while multi-antenna MA arrays require the joint optimization of individual antenna positions within confined movement regions. Consequently, the action-space dimensionality grows rapidly with the numbers of UAVs and antenna elements.

Motivated by the above challenges,  the main contributions of this work are summarized as follows:

\begin{itemize}

\item 
To enhance physical-layer security, we investigate a SAGIN-enabled secure downlink communication system in which UAVs select service links among satellite, aerial, and terrestrial networks while adjusting the positions of the MA array to fully exploit connectivity and spatial degrees of freedom for improved secrecy communication performance.
Specifically, we maximize the secrecy energy efficiency (SEE) of a UAV swarm by jointly optimizing the MA positions, UAV trajectories, and link selections, subject to UAV mobility, MA movement, and link connectivity constraints.

\item
To reduce CSI acquisition overhead, we propose a CKM-assisted MARL framework. The CKM is constructed offline from sparse channel measurements via Kriging interpolation and combined with satellite ephemeris information to enable CSI storage, retrieval, and reuse. Accordingly, each UAV proactively queries the CSI corresponding to its location and candidate MA configurations, and uses the retrieved information as interaction data for subsequent decision-making. This approach eliminates the additional overhead of real-time CSI acquisition during online decision-making.

\item
To address the strong coupling among link selection, UAV trajectory planning, and MA positioning, we construct a unified hybrid action space that enables a single policy to jointly generate the three types of decisions. An action-masking mechanism is employed to eliminate infeasible actions, thereby avoiding the suboptimality of stage-wise optimization. To overcome the high computational complexity of independently selecting the position of each antenna and the exponential growth in action dimensionality when jointly optimizing with other hybrid actions, we model the MA array using rigid-body kinematics, substantially reducing the action-space dimensionality. Furthermore, an individual-team hybrid reward mechanism is designed to coordinate local decisions with overall system performance: the individual reward improves the SEE of each UAV, whereas the team reward promotes swarm cooperation, thereby enhancing training efficiency and stability.

\end{itemize}
Simulation results validate the effectiveness of the proposed MA-positioning and link-selection degrees of freedom, as well as the contributions of the rigid-body modeling and individual-team hybrid reward mechanisms. Furthermore, the results demonstrate that the proposed framework achieves the optimal SEE of the UAV swarm while effectively reducing CSI acquisition and online inference overhead.

\textit{Organization:} The remainder of this paper is organized as follows. Section~\ref{System Model} presents the system model for cooperative secure communications by a UAV swarm in SAGIN. Section~\ref{Problem Formulation And Transformation} formulates the joint optimization problem. Section~\ref{Proposed CKM-Assisted HMAPPO Framework} introduces the proposed CKM-assisted MAPPO framework. Section~\ref{Simulation Results} presents the simulation results, and Section~\ref{Conclusion} concludes the paper.

\vspace{-3mm}

\section{System Model and Problem Formulation}\label{System Model}


In this section, we first present the network model for a SAGIN-assisted secure downlink communication system with a UAV swarm in Section~\ref{Network Model}. Section~\ref{Channel Model} develops the channel models for air-to-ground (A2G) and air-to-space (A2S) links, while Section~\ref{Signal Model} presents the corresponding signal models. 

\begin{figure}[!t]
	\centering
	\includegraphics[width=\columnwidth]{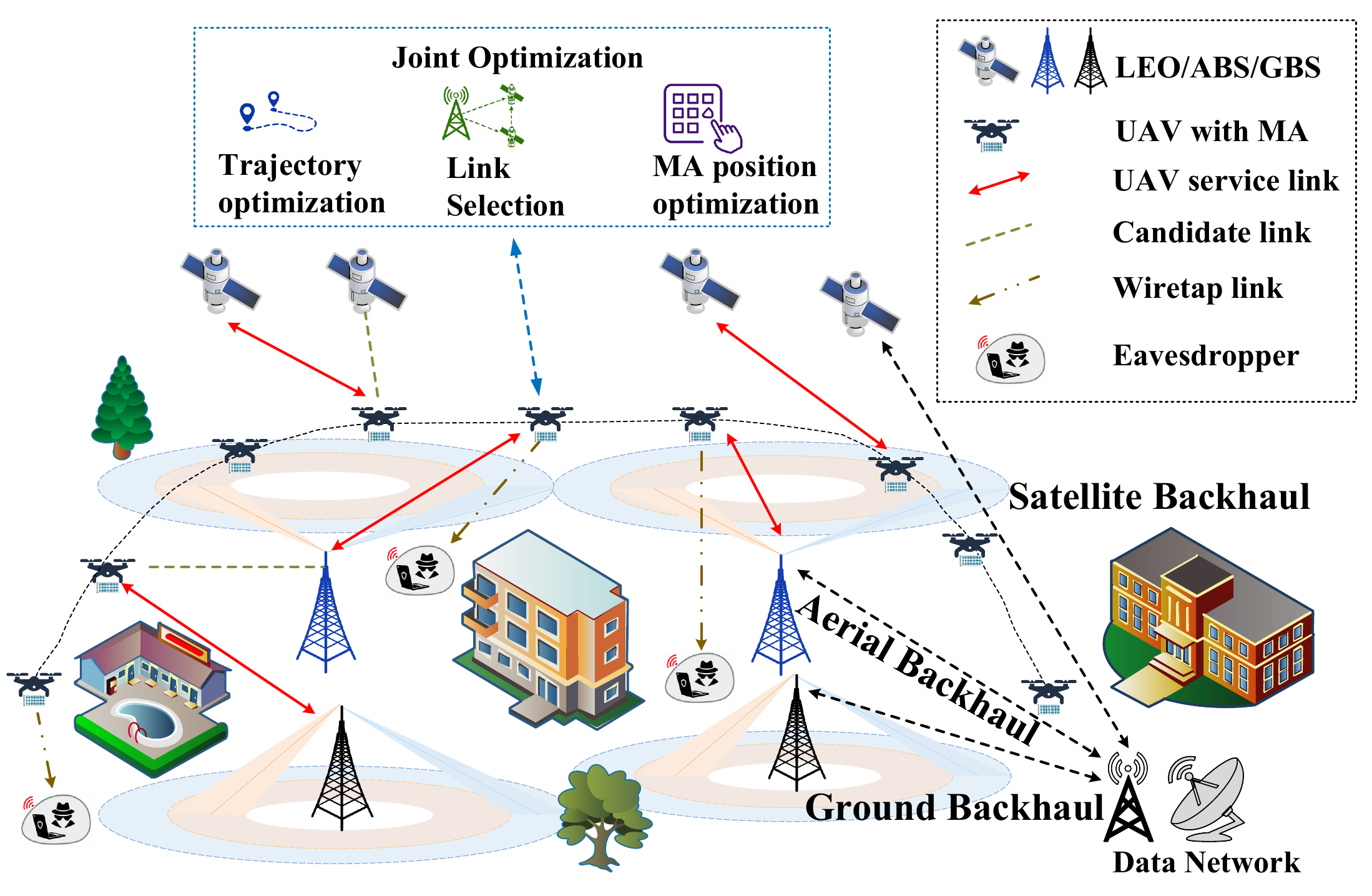}
	\caption{{Illustration of UAV secret communications in SAGIN with  LEO satellites, ABSs, and GBSs, under joint trajectory, MA, and link selection optimization.}}
	\label{fig1}
\end{figure}

\vspace{-3mm}

\subsection{Network Model}\label{Network Model}

We consider a SAGIN-assisted secure downlink communication system, in which low Earth orbit (LEO) satellites, aerial base stations (ABSs), and ground base stations (GBSs) provide downlink communication services to a UAV swarm. As illustrated in Fig.~\ref{fig1}, the UAV swarm cooperatively performs its tasks within a low-altitude service area jointly covered by space, aerial, and terrestrial networks.
Specifically, the TN consists of a GBS equipped with a uniform planar array (UPA) of $N^{\mathrm{GBS}}$ antenna elements. To ensure reliable and efficient communications in low-altitude airspace~\cite{kim2022non}, a dedicated ABS is deployed, equipped with a UPA of $N^{\mathrm{ABS}}$ antenna elements and an independently configured antenna uptilt angle. The SN comprises $S$ LEO satellites, each equipped with a multibeam antenna array of $N^{\mathrm{SN}}$ antenna elements and nadir-pointing beams fixed relative to the satellite body.
Within this architecture, the gateway station integrates the gateway functions of the TN, AN, and SN segments and interconnects the three network layers, thereby supporting efficient Internet content delivery~\cite{wang2024sustainable}. Meanwhile, a UAV swarm cooperatively performs secure communication tasks in low-altitude airspace, with each UAV equipped with an MA array comprising $M$ antenna elements~\cite{zhu2024movable}. We focus on two types of UAV-related communication links:
\begin{itemize}
    \item \textit{A2G links}: including GBS-UAV and ABS-UAV links, which connect a GBS {$b^{\mathrm{GBS}}$} and the $u$-th UAV, and an ABS {$b^{\mathrm{ABS}}$} and the $u$-th UAV, respectively.
    \item \textit{A2S links}: corresponding to LEO-UAV links, which connect a LEO satellite $b^{\mathrm{SN}}$ and the $u$-th UAV.
\end{itemize}
For A2G links, the corresponding BSs are assumed to be deployed at fixed altitudes, thereby providing coverage over specific service areas accessible to UAVs.
For A2S links, considering the high mobility of LEO satellites, we consider $S$ LEO satellites to ensure continuous coverage along the UAV flight trajectory, thereby guaranteeing that at least one satellite can maintain connectivity with the UAV at any time instant.
In this paper, we assume that each UAV can associate with only one serving BS at a time \cite{wang2024sustainable}.

\vspace{-4mm}

\subsection{Channel Model}\label{Channel Model}

We assume that each UAV is equipped with two functionally independent antenna subsystems~\cite{roste2026terminals}. The first is an $M$-element MA array mounted on a rectangular panel beneath the UAV, which establishes A2G links with GBSs and ABSs below the UAV~\cite{kim2026energy}. The second is a uniform linear array (ULA) mounted on the upper side of the UAV, which establishes A2S links with LEO satellites above the UAV~\cite{wang2024sustainable}. 
The two antenna subsystems are respectively oriented toward GBSs/ABSs below and LEO satellites above the UAV to accommodate the distinct spatial orientations and propagation characteristics of the two link types.

\subsubsection{MA-Enabled A2G Links}
\label{sec:ma-enabled-air-to-ground-links}

For the A2G link, this paper adopts a LoS-probability-based channel model~\cite{khuwaja2018survey}, where the dimensions of the transmit and receive regions are much smaller than the signal propagation distance. Suppose that the GBS or ABS is equipped with $K$ transmit antennas. {The position of the $m$-th MA element mounted on the $u$-th UAV is denoted by
$
\mathbf r_{u,m}
=
[x_{r,u,m},y_{r,u,m}]^T
\in\mathcal C_r,
\quad m=1,\ldots,M,
$}
where $\mathcal C_r$ denotes an $L_{\mathrm{MA}}\times L_{\mathrm{MA}}$ movable region~\cite{zhu2023movable}.
Let $\mathcal U$ denote the set of UAVs, and let
$
\mathcal M_u
=
\{\mathbf r_{u,1},\mathbf r_{u,2},\ldots,\mathbf r_{u,M}\}
$
denote the set of MA positions of the $u$-th UAV. The positions of all MAs mounted on the $u$-th UAV are stacked as
$
\tilde{\mathbf R}_u
=
[\mathbf r_{u,1},\mathbf r_{u,2},\ldots,\mathbf r_{u,M}]
\in\mathbb R^{2\times M},
$
where
\vspace{-1mm}
\begin{equation}
\lVert\mathbf r_{u,p}-\mathbf r_{u,l}\rVert_2\ge\widetilde d_{\min},\ \forall u\in\mathcal U,\ \forall p,l\in\{1,\ldots,M\},\ p\ne l.
\label{eq:problem_c7}
\end{equation}

The coordinates of the GBS (or ABS) and the UAV are respectively denoted by
$\mathbf q_B=[x_B,y_B,H_B]^T$ and
$\mathbf q_U=[x_U,y_U,H_U]^T$.
Accordingly, the link distance and the horizontal distance are given by
$
d=\|\mathbf q_U-\mathbf q_B\|_2,
\quad
d_h=\sqrt{(x_U-x_B)^2+(y_U-y_B)^2}.
$
The corresponding elevation angle is expressed as
$
\vartheta
=
\arctan\!\left(\frac{H_U-H_B}{d_h}\right).
$
{The elevation angle in degrees is defined as
$
\vartheta_{\mathrm{deg}}
=
\frac{180}{\pi}\vartheta.
$}
The LoS probability~\cite{khuwaja2018survey} is given by
$
P_{\mathrm{LoS}}^{\mathrm{A2G}}(\vartheta)
=
\frac{1}
{1+a\exp[-b(\vartheta_{\mathrm{deg}}-a)]},
$
where $a$ and $b$ are environment-dependent constants. The non-line-of-sight (NLoS) probability is
$
P_{\mathrm{NLoS}}^{\mathrm{A2G}}(\vartheta)
=
1-P_{\mathrm{LoS}}^{\mathrm{A2G}}(\vartheta).
$
Considering free-space path loss and additional LoS/NLoS losses, the average path loss is
\vspace{-1mm}
\begin{equation}
\mathrm{PL}(d,\vartheta)=20\log_{10}\!\left(\frac{4\pi d}{\lambda}\right)+P_{\mathrm{LoS}}^{\mathrm{A2G}}(\vartheta)\eta_{\mathrm{LoS}}+P_{\mathrm{NLoS}}^{\mathrm{A2G}}(\vartheta)\eta_{\mathrm{NLoS}}.
\end{equation}
where $\lambda$ is the carrier wavelength, and $\eta_{\mathrm{LoS}}$ and $\eta_{\mathrm{NLoS}}$ denote the additional losses of {LoS} and NLoS links, respectively. The large-scale channel power gain is
$
\beta(d,\vartheta)
=
10^{-\frac{\mathrm{PL}(d,\vartheta)}{10}}.
$
The small-scale MIMO channel is modeled as a geometric channel with $L$ effective paths. For the $l$-th path, let $\alpha_l$, $\mathbf u_{t,l}$, and $\mathbf u_{r,l}$ denote the complex gain, transmit direction, and receive direction, respectively.
The transmit array response of the fixed UPA is given by
$\mathbf a_t(\mathbf u_{t,l})=\left[e^{-j\frac{2\pi}{\lambda}\mathbf u_{t,l}^{T}\mathbf t_1},\ldots,e^{-j\frac{2\pi}{\lambda}\mathbf u_{t,l}^{T}\mathbf t_K}\right]^T,$
where $\mathbf t_K$ is the position of the $K$-th GBS (or ABS) antenna relative to the array reference point. For the movable receive array, {the two-dimensional position of the $m$-th MA mounted on the $u$-th UAV is extended to
$
\bar{\mathbf r}_{u,m}
=
[x_{r,u,m},y_{r,u,m},0]^T.
$}
Accordingly, the receive array response depending on the MA positions, namely the MA steering vector, is expressed as~\cite{ma2024mimo}
\vspace{-2mm}
\begin{equation}
{
\mathbf a_r(\tilde{\mathbf R}_u,\mathbf u_{r,l})
=
\left[
e^{-j\frac{2\pi}{\lambda}\mathbf u_{r,l}^{T}\bar{\mathbf r}_{u,1}},
\ldots,
e^{-j\frac{2\pi}{\lambda}\mathbf u_{r,l}^{T}\bar{\mathbf r}_{u,M}}
\right]^T.
}
\end{equation}
Unlike fixed arrays, each phase term in this steering vector can be continuously adjusted by the corresponding MA position $\bar{\mathbf r}_{u,m}$, which provides the additional spatial degrees of freedom introduced by MAs~\cite{ma2024mimo}. Thus, the MA-enabled A2G channel matrix is
\vspace{-2mm}
\begin{equation}
\mathbf H^{\mathrm{A2G}}(\tilde{\mathbf R}_u)
=
\sqrt{\beta(d,\vartheta)}
\sum_{l=1}^{L}
\alpha_l
\mathbf a_r(\tilde{\mathbf R}_u,\mathbf u_{r,l})
\mathbf a_t^{H}(\mathbf u_{t,l}).
\end{equation}
Since $\mathbf H^{\mathrm{A2G}}(\tilde{\mathbf R}_u)$ explicitly depends on $\tilde{\mathbf R}_u$, the UAV can improve the channel condition by optimizing $\tilde{\mathbf R}_u$~\cite{zhu2023movable}.

\subsubsection{A2S Links}
\label{sec:air-to-space-links}

Each UAV is assumed to be equipped with a $K$-element ULA mounted on its upper side. Accordingly, the A2S link is modeled as a {$K\times K$ channel}. Owing to the unobstructed LoS propagation path between the satellite and the UAV, and the fact that the transmission distance is much larger than the array apertures at both ends, a far-field single-path LoS channel model is adopted~\cite{wang2024sustainable}. The distance between the $s$-th satellite and the $u$-th UAV is expressed as
$
d_{s,u}
=
\left\|
\mathbf q_s-\mathbf q_u
\right\|_2,
$
where $\mathbf q_s$ and $\mathbf q_u$ denote the three-dimensional positions of the $s$-th satellite and the $u$-th UAV, respectively. The path loss of the A2S link~\cite{xiao2022antenna} is given by
$\mathrm{PL}_{s,u}^{\mathrm{A2S}}\!=\!20\log_{10}\!\left(\frac{4\pi d_{s,u}}{\lambda}\right)\!+\!\eta_{\mathrm{LoS}}.$
Accordingly, the large-scale channel power gain is
$
\beta_{s,u}^{\mathrm{A2S}}
\!=\!
10^{-\frac{\mathrm{PL}_{s,u}^{\mathrm{A2S}}}{10}}.
$
Define $\mathbf u_{s,u}=(\mathbf q_u-\mathbf q_s)/d_{s,u}$ as the unit direction vector from the $s$-th satellite to the $u$-th UAV. The satellite employs a $K_x\times K_y$ UPA, where $K=K_xK_y$.
Let $\mathbf t_{s,i}\in\mathbb R^3$ denote the position vector of the $i$-th antenna of the $s$-th satellite relative to the UPA reference point. The satellite UPA response is expressed as
$
\mathbf a_s(\mathbf u_{s,u})
\!=\!
\left[
e^{-j\frac{2\pi}{\lambda}\mathbf u_{s,u}^{T}\mathbf t_{s,1}},
\ldots,
e^{-j\frac{2\pi}{\lambda}\mathbf u_{s,u}^{T}\mathbf t_{s,K}}
\right]^T.
$
{Let $\mathbf p_{u,m}\in\mathbb R^3$ denote the position vector of the $m$-th element of the upper ULA of the $u$-th UAV relative to its reference point.} The UAV receive-array response is expressed as
$
{
\mathbf a_u(\mathbf u_{s,u})
\!=\!
\left[\!
e^{-j\frac{2\pi}{\lambda}\mathbf u_{s,u}^{T}\mathbf p_{u,1}},
\ldots,
e^{-j\frac{2\pi}{\lambda}\mathbf u_{s,u}^{T}\mathbf p_{u,K}}
\right]^T.
}\!
$
Consequently, the  channel matrix between the $s$-th satellite and the $u$-th UAV is expressed as~\cite{xiao2022antenna}
\vspace{-2mm}
\begin{equation}
\mathbf H_{s,u}^{\mathrm{A2S}}
\!=\!
\sqrt{\beta_{s,u}^{\mathrm{A2S}}}
\exp\!\left(
-j\frac{2\pi d_{s,u}}{\lambda}
\right)
{
\mathbf a_u(\mathbf u_{s,u})
\mathbf a_s^{H}(\mathbf u_{s,u})
}.
\end{equation}

\color{black}

\vspace{-4mm}

\subsection{Signal Model}\label{Signal Model}

 To characterize the time-varying nature of UAV mobility, a discrete time-slotted model is adopted \cite{zeng2021simultaneous}. Specifically, the total mission duration $T_s$ is equally divided into $N$ time slots, yielding $N+1$ discrete time instants indexed by $n=0,1,\ldots,N$. The duration of each time slot is denoted by $\delta_D$ and is defined as
$
    \delta_D=\frac{T_s}{N}.
$
It is assumed that the UAVs fly at a fixed altitude $H$.
Accordingly, the trajectory coordinates of the UAVs are subject to the following constraints:
\vspace{-2mm}
\begin{align}
    \mathbf q_u[0]=\mathbf q^{\text{ini}}_{u},\quad 
    \mathbf q_u[N]=\mathbf q^{\text{fin}}_{u},
    &\quad \forall u\in\mathcal{U}, \label{eq:mobility_init_final}\\
    \mathbf q_L \preceq \mathbf q_u[n] \preceq \mathbf q_U,
    &\quad \forall u\in\mathcal{U},\ \forall n, \label{eq:mobility_boundary}\\
    \left\|\mathbf q_u[n]-\mathbf q_v[n]\right\|_2 \ge d_{\min},
    &\quad \forall u,v\in\mathcal{U},\ \forall n, \label{eq:mobility_collision}
 \end{align}
where $\mathbf q_L\triangleq[x_L,y_L,H]^T$ and
$\mathbf q_U\triangleq[x_U,y_U,H]^T$ denote the lower and upper boundaries of the two-dimensional (2D) airspace, respectively.
\footnote{For simplicity, this paper focuses on two-dimensional trajectory optimization of UAVs, while three-dimensional  trajectory optimization is left for future work.}
 $\mathbf q^{\text{ini}}_{u}$ and $ \mathbf q^{\text{fin}}_{u}$ denote the initial and final positions of the $u$-th UAV, respectively.
 $\preceq$ denotes element-wise inequality and $d_{\min}$ is the minimum safety distance between any two UAVs.
 \eqref{eq:mobility_init_final} ensures that each UAV departs from its prescribed initial location and reaches its designated final location.  \eqref{eq:mobility_boundary} restricts the UAV trajectories to the specified flight region. Moreover, the collision-avoidance constraint in \eqref{eq:mobility_collision} requires a minimum safety distance of $d_{\min}$ to be maintained between any two UAVs, thereby preventing potential collisions.

Let $a_{u,b}[n]$ denote the link-selection indicator between the $u$-th UAV and the $b$-th BS at the $n$-th time slot, where the considered BSs comprise GBS, ABS, and LEO satellite. The set of all BSs is defined as
$\mathcal{B}=\mathcal{B}_{G}\cup\mathcal{B}_{A}\cup\mathcal{B}_{L},$
where $\mathcal{B}_{G}$, $\mathcal{B}_{A}$, and $\mathcal{B}_{L}$ denote the sets of GBSs, ABSs, and LEO satellites, respectively.  Accordingly, the link selections are subject to the following constraints:
\vspace{-2mm}
\begin{equation}
    a_{u,b}[n]\in\{0,1\},
    \quad \forall u\in\mathcal{U},\ \forall b\in\mathcal{B},\ \forall n,
    \label{eq:association_binary}
\end{equation}
\vspace{-2mm}
\begin{equation}
    \sum_{b\in\mathcal{B}} a_{u,b}[n]\leq 1,
    \quad \forall u\in\mathcal{U},\ \forall n,
    \label{eq:association_unique}
\end{equation}
\vspace{-2mm}
\begin{equation}
    \sum_{u=1}^{U} a_{u,b}[n]\leq C_b^{\max},
    \quad \forall b\in\mathcal{B},\ \forall n.
    \label{eq:association_capacity}
\end{equation}
where
 \eqref{eq:association_binary} specifies that $a_{u,b}[n]$ is a binary variable, where $a_{u,b}[n]=1$ indicates that the $u$-th UAV selects the $b$-th BS at time slot $n$, and $a_{u,b}[n]=0$ otherwise. 
\eqref{eq:association_unique} ensures that each UAV selects at most one BS at each time slot, while a zero summation indicates that the UAV selects no BS.
\eqref{eq:association_capacity} limits the number of UAVs served by each BS at each time slot to at most $C_b^{\max}$, thereby accounting for its limited communication capacity.
 Suppose that the $b$-th BS is
equipped with $N_b$ transmit antennas, while the $u$-th UAV is equipped
with $M$ receive antennas. 
Accordingly, the received signal after receive beamforming can be
expressed as
\vspace{-2mm}
\begingroup
\small
\begin{equation}
    \begin{aligned}
        & \textcolor{black}{y_{b,u}[n]}
        =
        \underbrace{
        \sqrt{P_{b,u}[n]}\,
        \mathbf v_{b,u}^{H}[n]
        \textcolor{black}{\mathbf H_{b,u}[n]}
        \mathbf w_{b,u}[n]
        s_{b,u}[n]
        }_{\text{desired signal}}
        \\
        &+\!\!
        \underbrace{
        \sum_{\substack{k\in\textcolor{black}{\mathcal{U}_b[n]}\\k\neq u}}\!
        \sqrt{P_{b,k}[n]}\,
        \mathbf v_{b,u}^{H}[n]
        \textcolor{black}{\mathbf H_{b,u}[n]}
        \mathbf w_{b,k}[n]
        s_{b,k}[n]
        }_{\text{intra-cell interference}}
        \!+\!
        \underbrace{
        \mathbf v_{b,u}^{H}[n]\mathbf z_{b,u}[n]
        }_{\text{noise}},
    \end{aligned}
    \label{eq:received_signal}
    \end{equation}
\endgroup
where $\mathcal{U}_b[n]\triangleq\{u\in\mathcal{U}\mid a_{u,b}[n]=1\}$ denotes the set of UAVs served by the $b$-th BS in time slot $n$, and $y_{b,u}[n]\in\mathbb{C}$ denotes the scalar signal after receive beamforming at the $u$-th UAV associated with the $b$-th BS. Moreover, $\mathbf{w}_{b,u}[n]\in\mathbb{C}^{N_b\times1}$ and $\mathbf{v}_{b,u}[n]\in\mathbb{C}^{M\times1}$ denote the transmit beamforming vector at the $b$-th BS for the $u$-th UAV and the receive beamforming vector at the $u$-th UAV, respectively, satisfying $\|\mathbf{v}_{b,u}[n]\|_2^2=1$ and $\|\mathbf{w}_{b,u}[n]\|_2^2\leq1$. Further, $P_{b,u}[n]$ denotes the transmit power from the $b$-th BS to the $u$-th UAV, $s_{b,u}[n]$ is the normalized information-bearing symbol satisfying $\mathbb{E}\{|s_{b,u}[n]|^2\}=1$, and $\mathbf{z}_{b,u}[n]\in\mathbb{C}^{M\times1}$ is the additive white Gaussian noise vector distributed as $\mathcal{CN}(\mathbf{0},\sigma_{b,u}^2\mathbf{I}_M)$.
Therefore, the received signal-to-interference-plus-noise ratio (SINR)
at the $u$-th UAV served by the $b$-th BS during the $n$-th time slot
can be expressed as
\vspace{-2mm}
\begin{equation}
    \gamma_{b,u}[n]
    =
    \frac{
        P_{b,u}[n]
        \left|
        \mathbf v_{b,u}^{H}[n]
        \textcolor{black}{\mathbf H_{b,u}[n]}
        \mathbf w_{b,u}[n]
        \right|^2
    }{
        \displaystyle
        \sum_{\substack{k\in\textcolor{black}{\mathcal{U}_b[n]}\\k\neq u}}
        P_{b,k}[n]
        \left|
        \mathbf v_{b,u}^{H}[n]
        \textcolor{black}{\mathbf H_{b,u}[n]}
        \mathbf w_{b,k}[n]
        \right|^2
        +
        \sigma_{b,u}^2
    }.
\label{eq:received_sinr}
\end{equation}

Based on the legitimate downlink model, we further characterize the wiretap link. The eavesdropping threat is assumed to exist only in the {GN band}, where Eve passively intercepts the downlink transmissions from the {GN} without actively interfering with legitimate communications~\cite{zhang2019securing}. Following the standard assumptions commonly adopted in physical-layer security studies, Eve is equipped with a single antenna, and her location, denoted by $\mathbf q_e$, is assumed to be known~\cite{zhang2019securing},\cite{kim2026energy}.
When the $b$-th GBS transmits data to the $u$-th UAV, the signals intended for the other UAVs in \textcolor{black}{$\mathcal{U}_b[n]$} act as multiuser interference at Eve. Accordingly, Eve's SINR for decoding the message intended for the $u$-th UAV is given by
\vspace{-1mm}
\begin{equation}
\gamma_{e,u,b}[n]
=
\frac{
P_{b,u}[n]\left|\mathbf h_{e,b}^{H}[n]\mathbf w_{b,u}[n]\right|^2
}{
\sum\limits_{\substack{k\in\textcolor{black}{\mathcal{U}_b[n]}\\ k\ne u}}
P_{b,k}[n]\left|\mathbf h_{e,b}^{H}[n]\mathbf w_{b,k}[n]\right|^2
+\sigma_e^2
},
\end{equation}
where $\mathbf h_{e,b}[n]\in\mathbb{C}^{N_b\times 1}$ denotes the known GBS-to-Eve channel.
\color{black}
Combining $\gamma_{e,u,b}[n]$ with the legitimate link SINR $\gamma_{b,u}[n]$ in \eqref{eq:received_sinr}, the achievable {GBS-link} secrecy rate of the $u$-th UAV associated with the $b$-th GBS at time slot $n$ is
\vspace{-1mm}
\begin{equation}
r_{u,b}^{\mathrm{sec,G}}[n]
\!=\!
\left[
\log_2\left(1+\gamma_{b,u}[n]\right)
\!-\!
\log_2\left(1+\gamma_{e,u,b}[n]\right)
\right]^+ \!,
 b\in\mathcal{B}_{G}.
\end{equation}
For {ABS-UAV} and LEO-UAV links, which are assumed to be free of eavesdropping, the secrecy rate reduces to the corresponding legitimate link rate. Thus, the secrecy rate over all candidate serving networks can be written uniformly as
\vspace{-1mm}
\begin{equation}
r_{u,b}^{\mathrm{sec}}[n]
=
\begin{cases}
\log_2\left(1+\gamma_{b,u}[n]\right),
&
b\in\mathcal{B}_{A}\cup\mathcal{B}_{L},
\\[1mm]
r_{u,b}^{\mathrm{sec,G}}[n],
&
b\in\mathcal{B}_{G}.
\end{cases}
\end{equation}
{Accordingly, the actual secrecy rate achieved by the $u$-th UAV  at time slot $n$, after accounting for its link selection, is defined as}
{\color{black}
\vspace{-2mm}
\begin{equation}
R_u^{\mathrm{sec}}[n]
=
\sum_{b\in\mathcal{B}}
a_{u,b}[n]r_{u,b}^{\mathrm{sec}}[n].
\label{eq:actual_secrecy_rate}
\end{equation}
}

\vspace{-2mm}
Given the limited onboard energy of the UAV, we further establish its energy consumption model. The total energy consumption \cite{zeng2019energy} comprises the propulsion, MA actuation, and communication energy expenditures, which are determined by $P_u$, $P_{\mathrm{MA}}$, and $P_{\mathrm{com}}$, respectively, along with their corresponding operating durations.
\color{black}
In this work, the $u$-th UAV is assumed to fly at a constant speed 
$\bar v_u$
throughout the entire mission. 
\footnote{
To simplify the trajectory design, each UAV is assumed to fly at a prescribed constant horizontal speed $\bar v_u$ \cite{zeng2021simultaneous}, which can be expressed as 
$
\bar v_u=\frac{1}{\delta_D}
\left\|\mathbf q_u[n+1]-\mathbf q_u[n]\right\|_2,
\quad \forall u\in\mathcal{U},\ \forall n.
\label{eq:mobility_speed}
$
}
Following the closed-form propulsion power model of rotary-wing UAVs 
\cite{zeng2019energy}, $P_{u}$
is expressed as
\vspace{-2mm}
\begin{equation}
\begin{aligned}
P_{u}
&=
P_0\left(1+\frac{3\bar v_u^2}{U_{\mathrm{tip}}^2}\right)
+
P_1\left(\sqrt{1+\frac{\bar v_u^4}{4v_0^4}}-\frac{\bar v_u^2}{2v_0^2}\right)^{\!\frac{1}{2}}
\\
&\quad+
\frac{1}{2}\,r_{\mathrm{drag}}\,\rho\,S\,A\,\bar v_u^3 ,
\end{aligned}
\end{equation}
where $P_0$ and $P_1$ denote the blade-profile power and the induced power in 
hovering status, respectively; $U_{\mathrm{tip}}$ is the rotor-blade tip speed; 
$v_0$ is the mean rotor-induced velocity in hovering; $r_{\mathrm{drag}}$ and 
$S$ represent the fuselage drag ratio and the rotor solidity, respectively;  
$A$ and $\rho$ denote the rotor-disk area and the air density, respectively.
\footnote{
Since the propulsion energy consumption dominates over that incurred by MA actuation and communication \cite{zeng2019energy}, only the propulsion power is retained in the energy model. Incorporating the energy consumption associated with MA actuation and communication is left for future work.
}

\vspace{-4mm}

    \section{Problem Formulation}\label{Problem Formulation And Transformation}

In this work, each UAV performs link selection among the satellite, aerial, and terrestrial networks while adjusting the positions of its MA array.
To fully exploit the connectivity and spatial degrees of freedom, we jointly optimize the MA positions $\tilde{\mathbf{R}}_u[n]$, UAV trajectories $\mathbf{q}u[n]$, and link-selection variables $a{u,b}[n]$ to maximize the SEE of the UAV swarm. The resulting optimization objective is formulated as follows:

    \vspace{-4mm}
    
    \begin{flalign}
        \mathcal{P}_1:
        & \max_{\{\mathbf{q}_u[n],\,\tilde{\mathbf{R}}_u[n],\,a_{u,b}[n]\}} \quad
        \frac{
        {\sum_{n=1}^{N}\sum_{\substack{u\in\mathcal{U}\\b\in\mathcal{B}}}
        a_{u,b}[n]r_{u,b}^{\mathrm{sec}}[n] \delta_D}
        }{
        \sum_{n=1}^{N}\sum_{u\in\mathcal{U}}
        {P_{u}^{\mathrm{tot}}[n]\delta_D}
        }
        && \label{eq:problem_obj}\\
        {\rm s.t.}\quad
        & \ \eqref{eq:problem_c7}, 
         \eqref{eq:mobility_init_final},\ \eqref{eq:mobility_boundary},\
          \eqref{eq:mobility_collision}, \eqref{eq:association_binary},\ \eqref{eq:association_unique},\
          \eqref{eq:association_capacity}. &&\notag
    \end{flalign}


It can be observed that $\mathcal{P}_1$ is a non-convex mixed-integer nonlinear programming (MINLP) problem, whose global optimum is generally difficult to obtain directly. Meanwhile, the evaluation of the secrecy rate $r_{u,b}^{\mathrm{sec}}[n]$ relies on instantaneous CSI, whereas CSI acquisition incurs substantial signaling overhead.
Moreover, the tight coupling among discrete link selection, continuous trajectory optimization, and MA position adjustment substantially enlarges the hybrid action space, which may impair sample efficiency and destabilize the training process. In particular, independently selecting the position of each antenna incurs considerable computational overhead and, when jointly optimized with other hybrid actions, can cause the cardinality of the joint action space to grow exponentially, thereby further increasing the solution complexity. The hard constraints on operational boundaries, safety separation, MA positions, and maximum association capacity, imposed by \eqref{eq:mobility_init_final}, \eqref{eq:mobility_boundary}, \eqref{eq:mobility_collision}, \eqref{eq:problem_c7}, and \eqref{eq:association_capacity}, must be strictly enforced throughout the mission; otherwise, the learned policy may generate infeasible or safety-critical trajectories.

\begin{figure*}[t]
    \centering
    \includegraphics[width=6in]{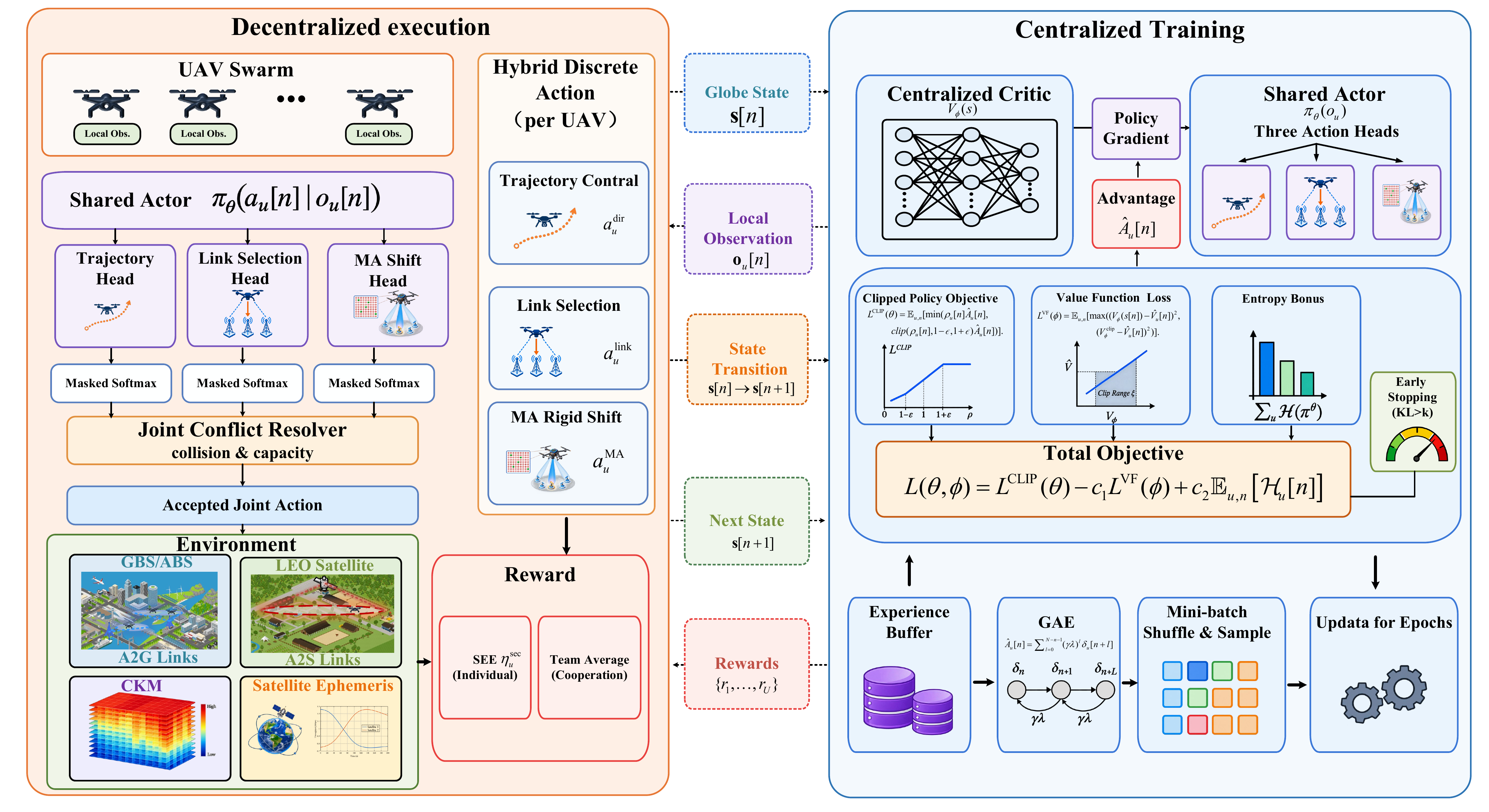}
    \captionsetup{justification=raggedright,singlelinecheck=false}
    \caption{\textcolor{black}{Proposed CKM-assisted MAPPO framework under centralized training with decentralized execution.}}
    \label{fig:mappo_framework}
\end{figure*}

To address these challenges, we develop three dedicated mechanisms: (i) exploiting the quasi-static nature of wireless channels in static-scatterer environments and their dependence on the UAV location and MA configuration, we construct a CKM, where the UAV location and candidate MA configuration jointly serve as the query key for CSI retrieval, thereby eliminating the need for real-time CSI acquisition;
(ii) we jointly model trajectory control, MA configuration, and link selection within a hybrid discrete action space, while modeling the MA array as a rigid body and controlling only its overall translation to reduce the action-space dimensionality; and (iii) we introduce an action-masking mechanism that excludes constraint-violating actions at each decision epoch, restricting policy optimization to the feasible action set. The detailed algorithmic framework is presented in the next section.

\vspace{-4mm}

\section{Proposed CKM-Assisted  MARL Framework}\label{Proposed CKM-Assisted HMAPPO Framework}

%

In this section, we propose a CKM-assisted MARL framework to solve the considered  problem. As illustrated in Fig.~\ref{fig:mappo_framework}, the framework employs an offline-constructed CKM and satellite ephemerides to characterize the training environment, enabling the agents to acquire CSI through data-driven interactions without requiring online interactions with the physical environment during training. 
To capture the strong coupling among link selection, UAV trajectory planning, and MA positioning, we formulate the three decisions as a unified hybrid discrete action space and employ action masking to ensure feasibility under hard constraints. To avoid the high computational cost of independent per-antenna position selection and the growth in action dimensionality caused by joint optimization, the MA array is parameterized using rigid-body kinematics, substantially reducing the action space. Finally, an individual-team hybrid reward is designed to improve the SEE of each UAV while promoting swarm cooperation.

\vspace{-2mm}

\subsection{CKM Construction for MARL Algorithm}
\label{3-1}


Before presenting the proposed CKM construction method, we briefly revisit the characteristics of wireless communication channels. In the considered SAGIN system, since the locations of the GBSs and ABSs are fixed, the variation in the MA-enabled MIMO channel $\mathbf{H}[t]$ is mainly determined by three factors: the UAV location $\mathbf{q}[t]$, the actual propagation environment $E[t]$ \cite{wu2023environment}, and the MA configuration $\tilde{\mathbf{R}}[t]$. Accordingly, a generic channel representation can be expressed as \cite{cheng2026channel}
\vspace{-1mm}
\begin{equation}
\color{black}
	\mathbf{H}[t] = f\bigl(\mathbf{q}[t], E[t], \tilde{\mathbf{R}}[t]\bigr),
\color{black}
\end{equation}
\color{black}
where \(f(\cdot,\cdot,\cdot)\) is an arbitrary function that maps from the {UAV location}, the propagation environment, and the MA positions to channel knowledge, and $\tilde{\mathbf{R}}[t]$ denotes the stacked MA position matrix defined earlier.
\color{black}
However, obtaining an accurate characterization of \(f(\cdot,\cdot,\cdot)\) in complex environments remains extremely challenging, primarily due to the difficulty of mathematically modeling the environment \(E[t]\) and capturing the intricate interactions between radio waves and their surroundings.
Fortunately, this issue can be effectively alleviated by leveraging the CKM concept in \cite{zeng2021toward}. Under the viewpoint widely adopted in the CKM literature \cite{wan2026channel}\cite{zeng2021toward}, the slowly varying geometry and large-scale statistical characteristics dominated by static structures such as buildings and walls can be approximated as piecewise quasi-static within a CKM validity interval, whereas blockage or reflection variations and the resulting channel transitions induced by mobile objects such as vehicles, trucks, and pedestrians are inevitable.

Building on this, we construct a Kriging-based CKM using a data-driven method \cite{li2023channel}. For the $b$-th BS, its CKM is denoted by $\mathcal{G}_b$ and consists of joint position--MA-configuration query keys and their corresponding channel-feature samples, i.e., $\mathcal{G}_b=\{(\boldsymbol{\xi}_i,\tilde{\mathbf z}_b(\boldsymbol{\xi}_i))\}_{i=1}^{N_{\mathrm{CKM}}}$. To account for the impact of the MA configuration on the channel, the query key is extended from a two-dimensional location to the joint position--configuration vector $\boldsymbol{\xi}_i=[\mathbf{x}_i^{\mathsf T},\operatorname{vec}(\tilde{\mathbf R}_i)^{\mathsf T}]^{\mathsf T}$, where $\mathbf{x}_i=[x_i,y_i]^{\mathsf T}\in\mathcal{D}\subseteq\mathbb{R}^2$ denotes the $i$-th sampling location, $\tilde{\mathbf R}_i$ is the MA configuration used for measurement, and $\tilde{\mathbf z}_b(\boldsymbol{\xi}_i)\in\mathbb{R}^{D_Z}$ denotes the corresponding measured channel-feature vector. 
For any unsampled joint query point $\boldsymbol{\xi}$, ordinary Kriging is employed to predict its channel feature from $\mathcal{G}_b$. Let $\{w_i^\ast\}_{i=1}^{N_{\mathrm{CKM}}}$ denote the interpolation weights determined under the minimum-variance unbiased criterion. The resulting prediction is given by\cite{li2023channel}\cite{cheng2026channel}
\vspace{-1.5mm}
\begin{equation}
    \mathbf{z}_b^{\mathrm{CKM}}(\boldsymbol{\xi})
    =
    \sum_{i=1}^{N_{\mathrm{CKM}}}
    w_i^\ast\tilde{\mathbf{z}}_b(\boldsymbol{\xi}_i).
    \label{eq:kriging_final}
\end{equation}

With the above method, once the CKM {$\mathbf z^{\mathrm{CKM}}(\boldsymbol{\xi})$} is obtained, the  CSI accumulated and stored over a long term by the ABS or GBS can be readily retrieved by inputting {both the current two-dimensional UAV location and the MA configuration}. 
Furthermore, by incorporating the SN link geometric information obtained from satellite ephemerides, the constructed CKM can efficiently provide CSI for the heterogeneous network at arbitrary UAV locations, MA configurations, and time instants.
It is worth noting that offline estimation may introduce uncertainties due to CKM reconstruction errors caused by sparse sampling, ephemeris prediction errors, and the mismatch between propagation models and practical wireless environments~\cite{zeng2021toward}.

\vspace{-4mm}

\subsection{MAPPO Algorithm for Joint UAV Link Selection, MA Position Optimization  and Trajectory Optimization}
\label{MAPPO Framework for Joint UAV Link Selection, MA Position Optimization and Trajectory Optimization}

To address the coupling among the decision variables, we propose a CKM-assisted multi-agent proximal policy optimization  with rigid-body shift design and hybrid rewards algorithm (MAPPO-RBHR-CKM).
Specifically, the considered problem is formulated as a cooperative multi-agent partially observable Markov decision process (POMDP), as illustrated in Fig.~\ref{fig:mappo_framework}, and is characterized by $\{\mathcal{S},\mathcal{A},\mathcal{O},\mathcal{R},\mathcal{P}\}$. Here, $\mathcal{S}$, $\mathcal{A}$, $\mathcal{O}$, $\mathcal{R}$, and $\mathcal{P}$ denote the global state space, joint action space, joint observation space, reward function, and state-transition probability, respectively. The POMDP elements are specified in the following.

\subsubsection{State Space}
The global state consists of the local state of each UAV and globally observable information. At time slot $n$, it is defined as
\vspace{-1.5mm}
\begin{equation}
\mathbf{s}[n]=\big[\{\mathbf{x}_u[n]\}_{u\in\mathcal{U}},c[n-1]\big],
\label{eq:uav_state_feature}
\end{equation}
where $\mathbf{x}_u[n]\triangleq[\mathbf q_u[n],\mathbf q_{u,\mathrm f},a_{u,b}[n-1],\mathbf z_u^{\mathrm{CKM}}[n],\tilde{\mathbf R}_u[n-1]]$ denotes the local state of the $u$-th UAV, and $c_u[n-1]$ denotes the locally observable network load.

\subsubsection{Observation Space}

During decentralized execution, UAV $u$ makes decisions based only on its local observation, defined at time slot $n$ as
\begin{equation}
\mathbf{o}_u[n]=\big[
\mathbf{x}_u[n],\mathcal{N}_u[n],c_u[n-1]
\big],\quad \forall u\in\mathcal{U},\ \forall n\in\mathcal{N},
\label{eq:uav_observation_feature}
\end{equation}
where $\mathcal{N}_u[n]$ represents the information of neighboring UAVs within its communication range. Following the centralized training and decentralized execution (CTDE) paradigm of MAPPO, the centralized critic takes the global state $\mathbf{s}[n]$ as input, whereas each decentralized actor relies only on $\mathbf{o}_u[n]$.

\subsubsection{Low-Dimensional Action-Masked Hybrid Action Space}\label{action}

To reduce the action-space complexity, UAV trajectory control, link selection, and MA configuration optimization are jointly formulated as a hybrid discrete action. The action of UAV $u$ at time slot $n$ is defined as
\begin{equation}
\mathbf{a}_u[n]=
\big[a_u^{\mathrm{dir}}[n],
a_u^{\mathrm{link}}[n],
a_u^{\mathrm{MA}}[n]\big],
\label{eq:joint_action}
\end{equation}
where $a_u^{\mathrm{dir}}[n]\in\{\theta_1,\ldots,\theta_{K_{\mathrm{dir}}}\}$ is the discrete heading action, and $a_u^{\mathrm{link}}[n]\in\mathcal{B}$ is the serving link selection action.

The conventional per-antenna independent selection scheme assigns each of the $M$ MAs independently to one of $N_{\mathrm{MA}}$ candidate locations, resulting in an MA action space with $N_{\mathrm{MA}}^M$ possible joint configurations. Consequently, the per-antenna formulation causes the action-space cardinality to grow exponentially with $M$; when jointly optimized with other hybrid actions, such as trajectory control and link selection, the resulting action space can become computationally prohibitive.
To address this issue, we model the entire MA array mounted on each UAV as a rigid body \cite{Mei2024Graph}, such that all MAs share a common translation vector. Let $\bar{\mathbf r}_m$ denote the nominal position of the $m$-th MA element. Specifically, the MA action is selected from $N_{\mathrm{sh}}$ candidate rigid-body translations, i.e.,
\begin{equation}
a_u^{\mathrm{MA}}[n]\in\{1,\ldots,N_{\mathrm{sh}}\},\quad
\mathbf r_{u,m}[n]=\bar{\mathbf r}_m+
\Delta\mathbf r_{a_u^{\mathrm{MA}}[n]}.
\label{eq:ma_shift_action}
\end{equation}
Accordingly, the MA action-space cardinality is reduced from $N_{\mathrm{MA}}^M$ under conventional per-antenna independent selection to $N_{\mathrm{sh}}$. 
Under the parameter setting considered in this work, the action-search complexity of the per-antenna formulation is $\mathcal{O}(49^4)$, whereas that of the proposed rigid-body translation formulation is reduced to $\mathcal{O}(9)$. 
This design effectively prevents the exponential growth of the MA action dimension, substantially reducing both the policy-network output dimension and the complexity of joint action exploration. 
The corresponding performance comparisons and ablation studies are further presented in Section~\ref{V-D} to quantitatively validate the effectiveness of the proposed rigid-body translation formulation in reducing the action-space complexity and improving joint optimization efficiency.

Furthermore, action masks are constructed for trajectory control, link selection, and MA configuration based on  constraints \eqref{eq:mobility_init_final}, \eqref{eq:mobility_boundary}, \eqref{eq:mobility_collision}, \eqref{eq:problem_c7}, \eqref{eq:association_capacity}, respectively. The resulting feasible sub-action spaces of UAV $u$ at time slot $n$ are given by
\vspace{-1.5mm}
\begin{equation}
\mathcal{A}_{u,n}^{\mathrm{dir}}
=
\left\{
a_u^{\mathrm{dir}}[n]\in\mathcal{A}^{\mathrm{dir}}
\,\middle|\,
\mathbf q_u^{+}[n]\in\mathcal Q,
\right\},
\label{eq:mask_dir}
\end{equation}
\begin{equation}
\mathcal{A}_{u,n}^{\mathrm{link}}
=
\left\{
b\in\mathcal{B}
\,\middle|\,
c_b[n-1]<C_b^{\max}
\right\},
\label{eq:mask_link}
\end{equation}
\begin{equation}
\mathcal{A}_{u,n}^{\mathrm{MA}}
=
\left\{
i\in\{1,\ldots,N_{\mathrm{sh}}\}
\,\middle|\,
\bar{\mathbf r}_m+\Delta\mathbf r_i\in\mathcal C_r,\ 
\forall m
\right\}.
\label{eq:mask_ma}
\end{equation}

Accordingly, the masked feasible hybrid action space is
\begin{equation}
\mathcal{A}_{u,n}
=
\mathcal{A}_{u,n}^{\mathrm{dir}}
\times
\mathcal{A}_{u,n}^{\mathrm{link}}
\times
\mathcal{A}_{u,n}^{\mathrm{MA}}.
\label{eq:hybrid_action_space}
\end{equation}

\subsubsection{Individual-Team Collaborative Reward Function}

Different from conventional designs that rely solely on either individual or team rewards, this paper proposes an individual-team collaborative reward function to simultaneously balance local performance optimization and system-level cooperation for UAVs. Specifically, the individual reward of UAV $u$ at the $n$-th time slot is defined as
\vspace{-1.5mm}
\begin{equation}
r_{u,n} = \beta r_{u,n}^{\mathrm{ind}} + (1-\beta) r_n^{\mathrm{team}},
\end{equation}
where $r_n^{\mathrm{team}}$ is defined as
$
r_n^{\mathrm{team}} = \frac{1}{U}\sum_{u=1}^{U} r_{u,n}^{\mathrm{ind}}
$.
\color{black}
Unlike conventional reward mechanisms that mainly rely on instantaneous communication performance and a fixed per-step penalty, we take the SEE as the core component of the reward to guide UAV trajectory optimization during flight. Specifically, the SEE term of the reward at time slot \(n\) is defined as
\vspace{-1.5mm}
{\color{black}
\begin{equation}
    r^{\mathrm{ind}}_{u,n}
    = \eta^{\mathrm{sec}}_{u,n}
    = \frac{R_u^{\mathrm{sec}}[n]\delta_D}
    {{P_{u}^{\mathrm{tot}}[n]}\delta_D}
    = \frac{\sum_{b\in\mathcal B}a_{u,b}[n]r_{u,b}^{\mathrm{sec}}[n]}
    {{P_u^{\mathrm{tot}}[n]}},
    \label{eq:reward_sec}
\end{equation}
}

\begin{algorithm}[t]
\caption{MAPPO-RBHR-CKM}
\label{alg:mappo}
\begin{algorithmic}[1]
\State \textbf{Input:} CKM, ephemerides, $\gamma$, $\lambda$, $\epsilon$, $\xi$, $\kappa$
\State \textbf{Initialize:} $\pi_{\bm\theta}$, $V_{\bm\phi}$, $\mathcal{D}\leftarrow\varnothing$
\For{episode $=1$ to $E$}
    \State Reset environment and obtain $\mathbf{s}[0]$
    \For{$n=0$ to $N-1$}
        \For{each $u\in\mathcal{U}$ \textbf{in parallel}}
            \State Query CKM and form $\mathbf{o}_u[n]$
            \State Mask $\mathcal{A}_{u,n}^{\mathrm{dir}}$, $\mathcal{A}_{u,n}^{\mathrm{link}}$, $\mathcal{A}_{u,n}^{\mathrm{MA}}$
            \State Sample $\tilde a_u[n]\sim\pi_{\bm\theta}(\cdot\mid\mathbf{o}_u[n])$
        \EndFor
        \State Resolve conflicts and obtain $\mathbf{a}[n]$
        \State Execute $\mathbf{a}[n]$; obtain $\mathbf{s}[n+1]$ and $r_{u,n}$
        \State Store $\{\mathbf{o}_u[n],\mathbf{s}[n],a_u[n],r_{u,n}\}$ in $\mathcal{D}$
    \EndFor
    \State Compute $\hat{A}_u[n]$ and $\hat{V}_u[n]$ by~\eqref{eq:gae}
    \State $\bm\theta_{\mathrm{old}}\leftarrow\bm\theta$, $\bm\phi_{\mathrm{old}}\leftarrow\bm\phi$
    \For{epoch $=1$ to $E_{\mathrm{ppo}}$}
        \For{each mini-batch from $\mathcal{D}$}
            \State Update $\bm\theta$ by~\eqref{eq:clip},~\eqref{eq:total}
            \State Update $\bm\phi$ by minimizing~\eqref{eq:vf}
        \EndFor
        \State \textbf{if} $\mathrm{KL}(\pi_{\bm\theta_{\mathrm{old}}}\parallel\pi_{\bm\theta})>\kappa$ \textbf{then break}
    \EndFor
    \State $\mathcal{D}\leftarrow\varnothing$
\EndFor
\State \textbf{Output:} $\pi_{\bm\theta}$
\end{algorithmic}
\end{algorithm}

\subsubsection{MAPPO Algorithm}
\label{subsec:lagrangian_cmapppo}

Based on the above POMDP formulation, MAPPO is adopted under the CTDE paradigm to solve the joint optimization problem. Since all UAVs are homogeneous agents, their actors share the policy parameters $\pi_{\bm\theta}$, thereby improving scalability and preserving permutation invariance. During training, the centralized critic $V_{\bm\phi}$ performs value estimation based on the global state $\mathbf{s}[n]$ to facilitate policy learning. During execution, the critic is removed, and each UAV independently makes decisions based only on its local observation $\mathbf{o}_u[n]$. Therefore, the decision-making process remains partially observable during decentralized execution, whereas global information is used exclusively for policy learning during centralized training, as shown in Fig.~\ref{fig:mappo_framework}.

The hybrid action $a_u[n]=[a_u^{\mathrm{dir}}[n],a_u^{\mathrm{link}}[n],a_u^{\mathrm{MA}}[n]]$ consists of three conditionally independent sub-actions. Hence, given the observation, the policy factorizes as
\vspace{-1mm}
\begin{equation}
    \pi_{\bm\theta}\!\left(a_u[n]\mid \mathbf{o}_u[n]\right)
    =\!\!\!\prod_{x\in\{\mathrm{dir},\mathrm{link},\mathrm{MA}\}}\!\!\!
    \pi_{\bm\theta}^{x}\!\left(a_u^{x}[n]\mid \mathbf{o}_u[n]\right),
    \label{eq:policy_factorize}
\end{equation}
{where each sub-policy head is realized by a masked categorical distribution \cite{huang2022invalid}. Specifically, let $\mathbf{z}_u^{x}[n]$ denote the logits produced by head $x$. Local infeasible actions receive logits of $-\infty$ before softmax; the joint resolver then handles simultaneous collision and residual-capacity conflicts. Taking the trajectory head as an example, the masked distribution is}
\vspace{-1mm}
\begin{equation} \pi_{\bm\theta}^{\mathrm{dir}}\!\left(a\mid\mathbf{o}_u[n]\right)
    =\frac{\exp\!\big(z_{u,a}^{\mathrm{dir}}[n]\big)\,
            \mathbbm{1}\!\left\{a\in\mathcal{A}_{u,n}^{\mathrm{dir}}\right\}}
           {\sum_{a'\in\mathcal{A}_{u,n}^{\mathrm{dir}}}
            \exp\!\big(z_{u,a'}^{\mathrm{dir}}[n]\big)},
    \label{eq:masked_softmax}
\end{equation}
The link and MA heads adopt the same masked form over $\mathcal{A}_{u,n}^{\mathrm{link}}$ and $\mathcal{A}_{u,n}^{\mathrm{MA}}$, respectively. Owing to the factorization in~\eqref{eq:policy_factorize}, the joint log-probability and entropy of the hybrid action are simply the sums over the three heads.
{With the centralized critic $V_{\bm\phi}(\mathbf{s}[n])$, the temporal-difference (TD) error is $\delta_u[n]=r_{u,n}+\gamma V_{\bm\phi}(\mathbf{s}[n{+}1])-V_{\bm\phi}(\mathbf{s}[n])$.} The advantage is then estimated by the generalized advantage estimator (GAE) to trade off bias and variance:
\vspace{-1mm}
\begin{equation}
    \hat{A}_u[n]=\sum_{l=0}^{N-n-1}(\gamma\lambda)^{l}\,\delta_u[n{+}l],
    \label{eq:gae}
\end{equation}
where $\lambda\in[0,1]$ is the GAE decay factor and $N$ is the number of slots per episode. The corresponding value target is {$\hat{V}_u[n]=\hat{A}_u[n]+V_{\bm\phi_{\mathrm{old}}}(\mathbf{s}[n])$}.
Let $\rho_u[n](\bm\theta)=\dfrac{\pi_{\bm\theta}(a_u[n]\mid\mathbf{o}_u[n])}{\pi_{\bm\theta_{\mathrm{old}}}(a_u[n]\mid\mathbf{o}_u[n])}$ denote the probability ratio between the new and old policies. 
{Proximal policy optimization (PPO)} maximizes the clipped surrogate objective to prevent excessively large policy updates:
\vspace{-1mm}
\begin{equation}
    \begin{aligned}
        L^{\mathrm{CLIP}}(\bm\theta)
        &= \mathbb{E}_{u,n}\Big[
            \min\big(\rho_u[n]\hat{A}_u[n], \\
        &\quad\ \mathrm{clip}(\rho_u[n],1{-}\epsilon,1{+}\epsilon)\hat{A}_u[n]\big)
        \Big],
    \end{aligned}
    \label{eq:clip}
\end{equation}
where $\epsilon$ is the clipping threshold and the advantages are normalized within each mini-batch to stabilize learning.
The critic is updated by minimizing a clipped value loss to suppress drastic fluctuations of the value estimate:
\vspace{-1mm}
\begin{equation}
    L^{\mathrm{VF}}(\bm\phi)
    =\mathbb{E}_{u,n}\!\Big[\!
        \max\!\big(
            (V_{\bm\phi}({\mathbf{s}[n]}){-}\hat{V}_u[n])^2,
            (V_{\bm\phi}^{\mathrm{clip}}{-}\hat{V}_u[n])^2
        \big)\!\Big],
    \label{eq:vf}
\end{equation}
where {$V_{\bm\phi}^{\mathrm{clip}}\!=\!V_{\bm\phi_{\mathrm{old}}}(\mathbf{s}[n])+\operatorname{clip}\!\big(V_{\bm\phi}(\mathbf{s}[n])-V_{\bm\phi_{\mathrm{old}}}(\mathbf{s}[n]),-\xi,\xi\big)$} and $\xi$ is the value-clipping threshold.
To encourage exploration and avoid premature convergence, the entropy of the hybrid policy, given by the sum over the three heads
$\mathcal{H}_u[n]=\sum_{x}\mathcal{H}\!\big[\pi_{\bm\theta}^{x}(\cdot\mid\mathbf{o}_u[n])\big]$,
is incorporated as a regularizer. Combining~\eqref{eq:clip},~\eqref{eq:vf} and the entropy term, the overall objective is
\vspace{-1mm}
\begin{equation}
    L(\bm\theta,\bm\phi)=
    L^{\mathrm{CLIP}}(\bm\theta)
    -c_1 L^{\mathrm{VF}}(\bm\phi)
    +c_2\,\mathbb{E}_{u,n}\!\left[\mathcal{H}_u[n]\right],
    \label{eq:total}
\end{equation}
where $c_1$ and $c_2$ are the weighting coefficients of the value loss and the entropy regularizer, respectively, and both networks are optimized by the Adam optimizer.

\begin{table}[t]
    \centering
    \footnotesize
    \captionsetup{font=scriptsize}
    \caption{System Simulation Parameters}
    \label{tab:System_params}
    \setlength{\tabcolsep}{2pt}
    \renewcommand{\arraystretch}{1}

    \begin{tabular}{p{0.30\columnwidth} p{0.20\columnwidth} p{0.20\columnwidth} p{0.20\columnwidth}}
        \toprule
        \textbf{System Parameter} & \textbf{GBS}\cite{zeng2021simultaneous} & \textbf{ABS}\cite{kim2022non} & \textbf{LEO}\cite{wang2025multi} \\
        \midrule
        Carrier frequency
        & $6.7$~GHz & $4.7$~GHz & $2.185$~GHz \\
        Transmit power
        & $46$~dBm & $46$~dBm & $34$~dBW \\
        Antenna array
        & $4\times4$ UPA & $4\times4$ UPA & $8\times8$ UPA \\
        Element spacing
        & $\lambda/2$ & $\lambda/2$ & $\lambda/2$ \\
        Beam direction
        & Downtilted & Uptilted & Vertical \\
        Height
        & $25$~m & $25$~m & $550$~km \\
        Connectivity capacity
        & $8$ & $8$ & $8$ \\
        \bottomrule
    \end{tabular}
\end{table}

\begin{table}[t]
    \centering
    \footnotesize
    \captionsetup{font=scriptsize}
    \caption{UAV Simulation Parameters}
    \label{tab:uav_params}
    \setlength{\tabcolsep}{2pt}
    \renewcommand{\arraystretch}{1}

    \begin{tabular}{p{0.48\columnwidth} p{0.48\columnwidth}}
        \toprule
        \textbf{UAV Parameter} & \textbf{Value} \\
        \midrule
        Number of UAVs
        & $20$ \\
        Flight altitude
        & $H=100$~m \\
        Maximum flight speed
        & $\bar{v}_u=20$~m/s \\
        Bandwidth per UAV
    & $1$~MHz \\
        Minimum UAV spacing
        & $d_{\min}=3$~m \\
        MA array
        & $4$ movable antennas \\
        Minimum MA spacing
        & $\widetilde{d}_{\min}=\lambda/2$ \\
         MA movable region \cite{zhu2023movable}
        & $[-D,D]\times[-D,D]$, $D=4\lambda$ \\
        Blade profile power
        & $P_0=79.86$~W \\
        Induced power \cite{zeng2019energy}
        & $P_1=88.63$~W \\
        Rotor-blade tip speed \cite{zeng2019energy}
        & $U_{\mathrm{tip}}=120$~m/s \\
        Mean rotor-induced velocity
        & $v_0=4.03$~m/s \\
        Fuselage drag ratio
        & $r_{\mathrm{drag}}=0.6$ \\
        Atmospheric density
        & $\rho=1.225$~kg/m\textsuperscript{3} \\
        Rotor solidity \cite{zeng2019energy}
        & $S=0.05$ \\
        Rotor disk area \cite{zeng2019energy}
        & $A_{\mathrm{rot}}=0.503$~m\textsuperscript{2} \\
        \bottomrule
    \end{tabular}
\end{table}

\begin{table}[t]
	\centering
	\footnotesize
	\captionsetup{font=scriptsize}
	\caption{Network Parameters}
	\label{tab:network_params}
	\setlength{\tabcolsep}{3pt}
	\begin{tabular}{p{0.58\columnwidth} p{0.34\columnwidth}}
		\toprule
		\textbf{Parameter} & \textbf{Value} \\
		\midrule
		Actor hidden layers & $128,\,128$ \\
		Critic hidden layers & $256,\,256,\,256$ \\
		Activation function & Tanh \\
		Optimizer & Adam \\
		Actor/critic learning rate & $1\times10^{-4}$ \\
		Discount factor & $0.99$ \\
		GAE parameter & $0.95$ \\
		PPO clipping parameter & $0.2$ \\
		PPO epochs per update & $E_{\mathrm{ppo}}=10$ \\
		Mini-batch size & $256$ \\
		Hybrid-reward weight & $\beta=0.5$ \\
		Value/entropy weights & $c_1=0.5,\ c_2=0.01$ \\
		Value clip / target KL & $\xi=0.2,\ \kappa=0.015$ \\
		No. of rigid-body shifts & $9$ \\
		Max. time slots per episode & $40$ \\
		No. of training episodes & $10000$ \\
		No. of independent seeds & $10$ \\
		\bottomrule
	\end{tabular}
	\vspace{-2mm}
\end{table}

\vspace{-2mm}

\section{Numerical Results}\label{Simulation Results}

This section first presents the simulation setup and parameter configurations in Section~\ref{V-A}, followed by the benchmark schemes in Section~\ref{V-B}. Section~\ref{V-C} evaluates the secrecy energy efficiency performance of the proposed algorithm, while Section~\ref{V-D} presents ablation studies to validate the effectiveness of the two proposed mechanisms and the considered scenario design. Finally, Section~\ref{V-E} analyzes the computational complexity and signaling overhead.

\vspace{-3mm}

\subsection{Simulation Setup}\label{V-A}

\subsubsection{Communication Parameter Settings}


We consider a UAV swarm comprising $20$ UAVs, where each UAV departs from a predefined initial location and flies at a fixed altitude $H$ toward its designated destination while performing secure communication within a $500\,\mathrm{m}\times500\,\mathrm{m}$ low-altitude area. The considered heterogeneous network consists of one GBS, one ABS, and $2$ LEO satellites, with the corresponding system parameters listed in Table~\ref{tab:System_params}. Full frequency reuse is adopted, such that each UAV is allocated $1$~MHz bandwidth for its serving link, while concurrent transmissions over the same spectrum result in co-channel interference. The UAV-related parameters are provided in Table~\ref{tab:uav_params}. For the A2G channel model, the urban-environment parameters are set as $a=9.61$, $b=0.16$, $\eta_{\mathrm{LoS}}=1$~dB, and $\eta_{\mathrm{NLoS}}=20$~dB, with $L=3$ geometric propagation paths \cite{zhan2022energy}.

\subsubsection{Neural Network and Training Parameter Settings}

For the proposed MAPPO-RBHR-CKM algorithm, all UAVs share an actor network with two fully connected hidden layers of $128$ neurons each. Separate output heads are used for flight direction, link selection, and MA position selection. For MA control, the entire $2\times2$ array selects one of $9$ candidate rigid-body shifts.
{A permutation-invariant centralized critic uses three fully connected layers of $256$ neurons and mean pooling to aggregate agent features. Both networks use Adam with a learning rate of $1\times10^{-4}$. Each episode contains $40$ slots and training lasts $10000$ episodes. The detailed parameters are summarized in Table~\ref{tab:network_params}.}

\begin{figure*}[t]
    \centering
    
    \begin{subfigure}[t]{0.31\textwidth}
        \centering
        \includegraphics[width=\linewidth]{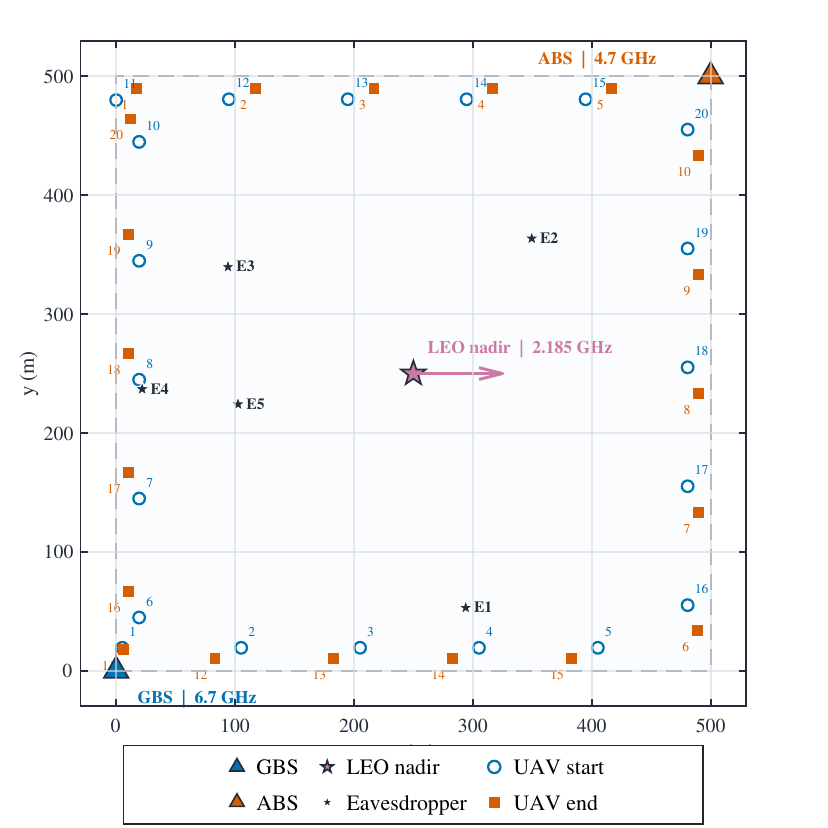}
        \caption{Illustration of the SAGIN simulation scenario.}
        \label{fig1a}
    \end{subfigure}\hfill
    \begin{subfigure}[t]{0.31\textwidth}
        \centering
        \includegraphics[width=\linewidth]{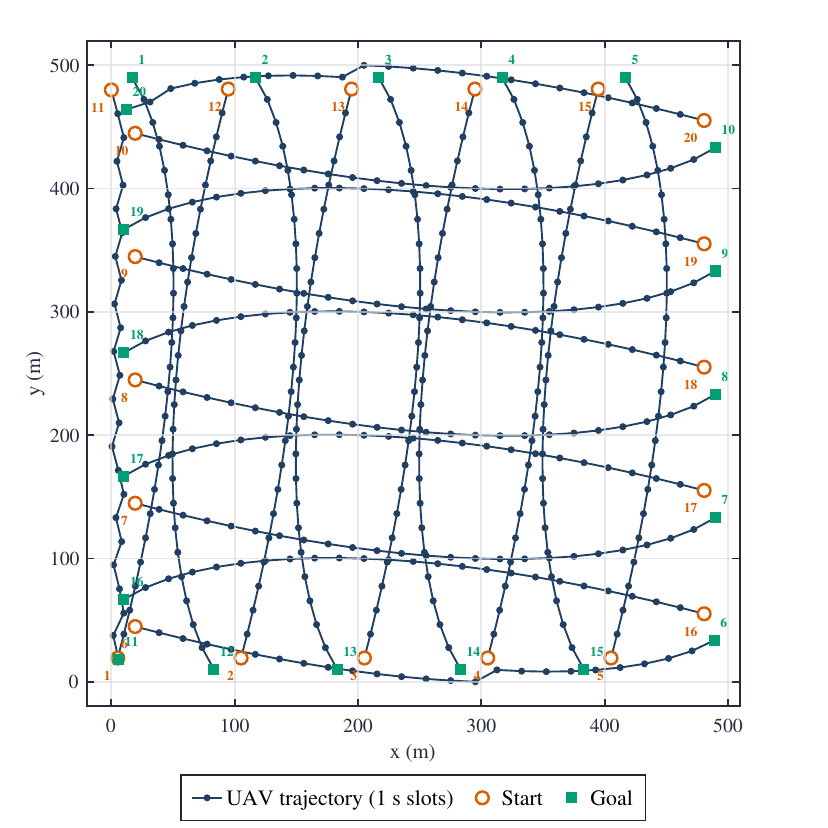}
        \caption{Illustration of the trajectories of a swarm of UAVs.}
        \label{fig1b}
    \end{subfigure}\hfill
    \begin{subfigure}[t]{0.30\textwidth}
        \centering
        \includegraphics[width=\linewidth]{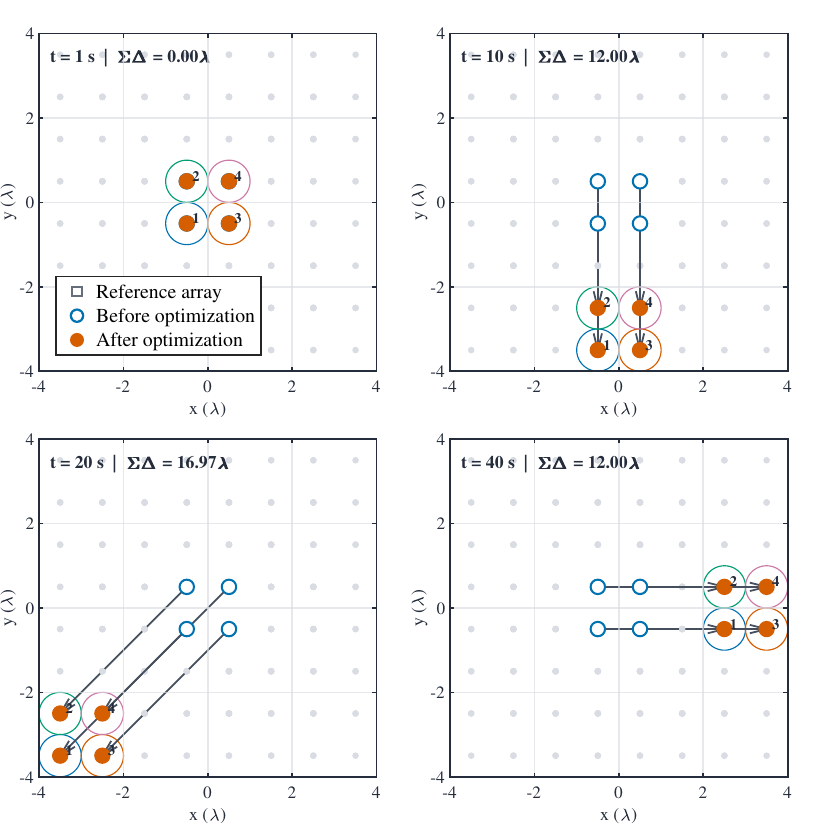}
        \caption{Illustration of MA positions before and after optimization.}
        \label{fig1c}
    \end{subfigure}
    
    {\captionsetup{justification=raggedright,singlelinecheck=false}
        \caption{Visualization of proposed algorithm.}
        \label{fig_1UAV}}
\end{figure*}

\begin{figure}[t]
	\centering
	\includegraphics[width=\columnwidth]{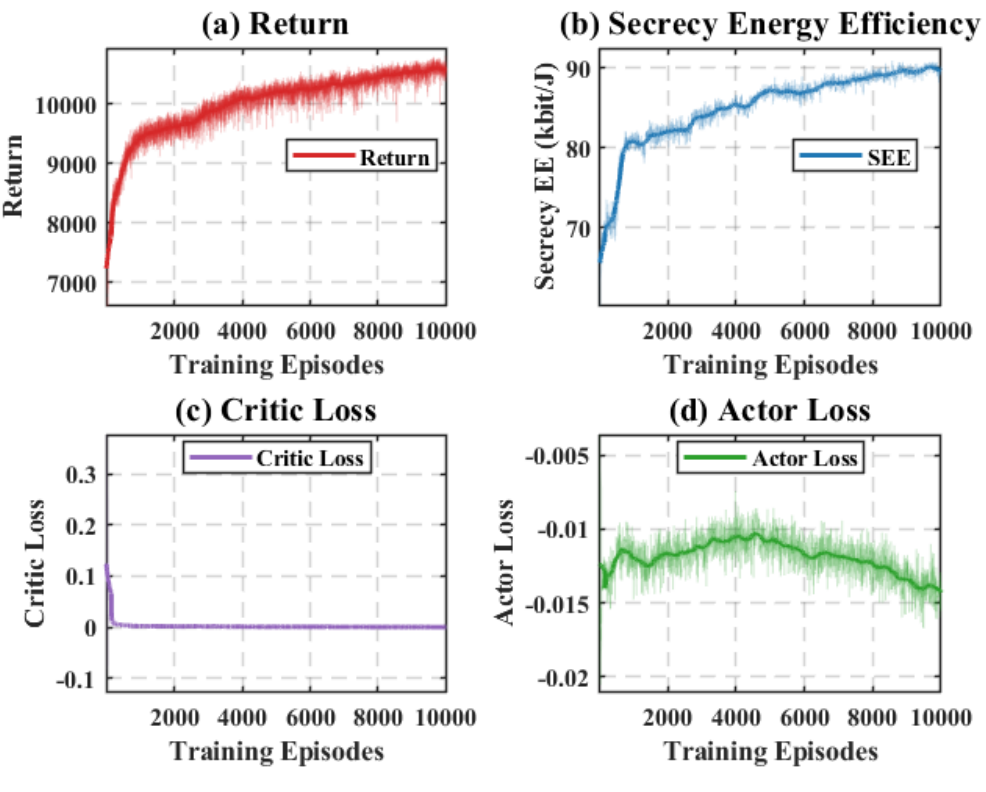}
	\caption{Illustration of the training performance of the proposed algorithm.}
	\label{fig3}
\end{figure}

\begin{figure}[t]
	\centering
	\includegraphics[width=\columnwidth]{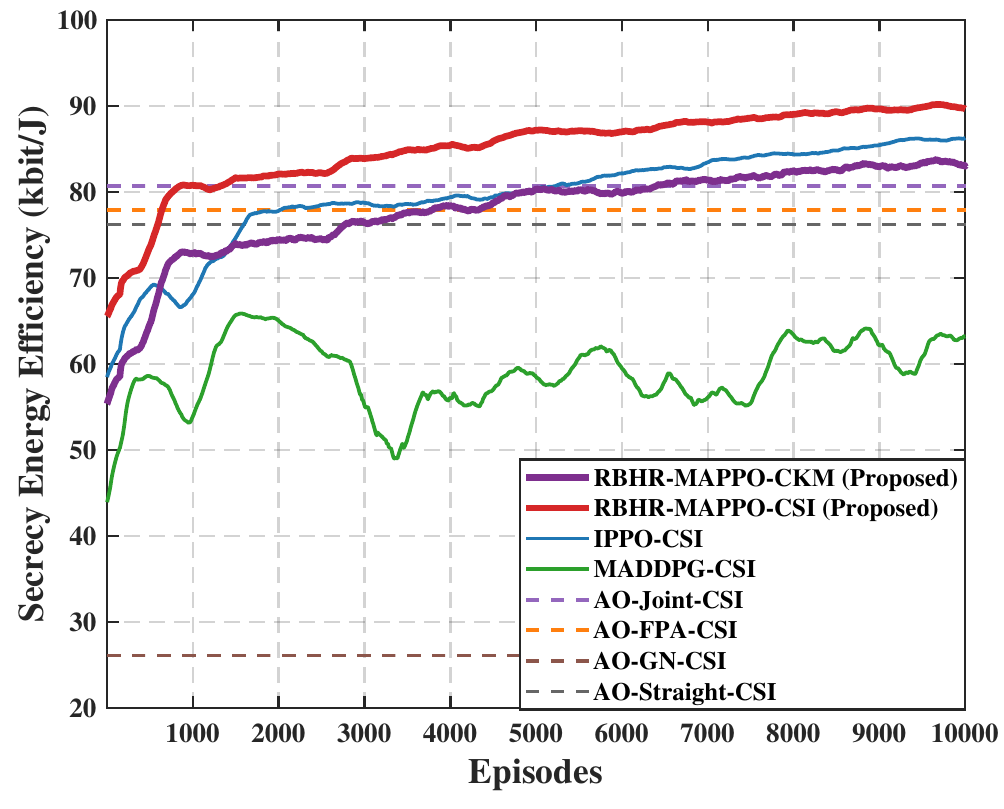}
	\caption{{SEE comparison of the proposed algorithms and baseline schemes during training.}}
	\label{fig4}
\end{figure}

\vspace{-5mm}

\subsection{Benchmarks}\label{V-B}

{To evaluate the proposed algorithms, we compare it with established learning baselines and corresponding convex-optimization-based alternating optimization (AO) methods. These AO methods follow the alternating-optimization principles in~\cite{zhu2024movable}\cite{shao2025network} and are adapted to the joint optimization problem considered in this work, rather than being directly adopted from those references:}

\begin{itemize}
	\item \textbf{AO-Joint-CSI}: An AO variant that jointly updates UAV trajectories, link selections, and MA positions.
	
	\item \textbf{AO-FPA-CSI}: An AO variant with FPAs, jointly optimizing trajectories and link selections.
	
	\item \textbf{AO-GN-CSI}: An AO variant with a fixed GN link selection, jointly optimizing trajectories and MA positions.
	
	\item \textbf{AO-Straight-CSI}: An AO variant with predefined straight-line trajectories, jointly optimizing link selections and MA positions.

	\item \textbf{IPPO-CSI}: An independent proximal policy optimization (IPPO) baseline \cite{deWitt2020IPPO} using perfect instantaneous CSI, where each UAV independently optimizes its trajectory, link selection, and MA positions.
	
	\item \textbf{MADDPG-CSI}: A multi-agent deep deterministic policy gradient (MADDPG) baseline \cite{lowe2017maddpg} using perfect instantaneous CSI to jointly optimize UAV trajectories, link selections, and MA positions.
	
\item \textbf{Proposed MAPPO-RBHR-CKM/CSI}: The proposed MAPPO-RBHR framework takes either CKM information or perfect instantaneous CSI as input to jointly optimize UAV trajectories, link selections, and MA positions.

\end{itemize}

\begin{figure*}[t]
    \centering
    
    \begin{subfigure}[t]{0.32\textwidth}
        \centering
        \includegraphics[width=\linewidth]{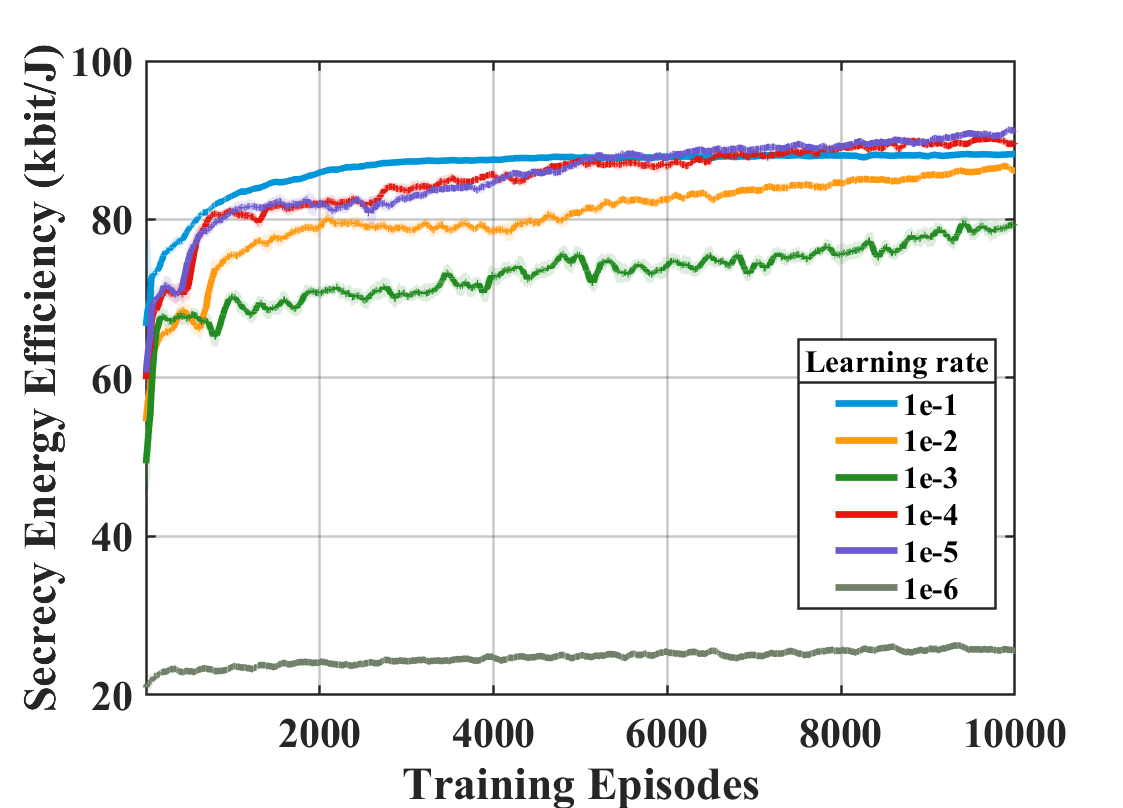}
        \caption{Impact of learning rate on algorithm performance.}
        \label{fig5a}
    \end{subfigure}\hfill
    \begin{subfigure}[t]{0.32\textwidth}
        \centering
        \includegraphics[width=\linewidth]{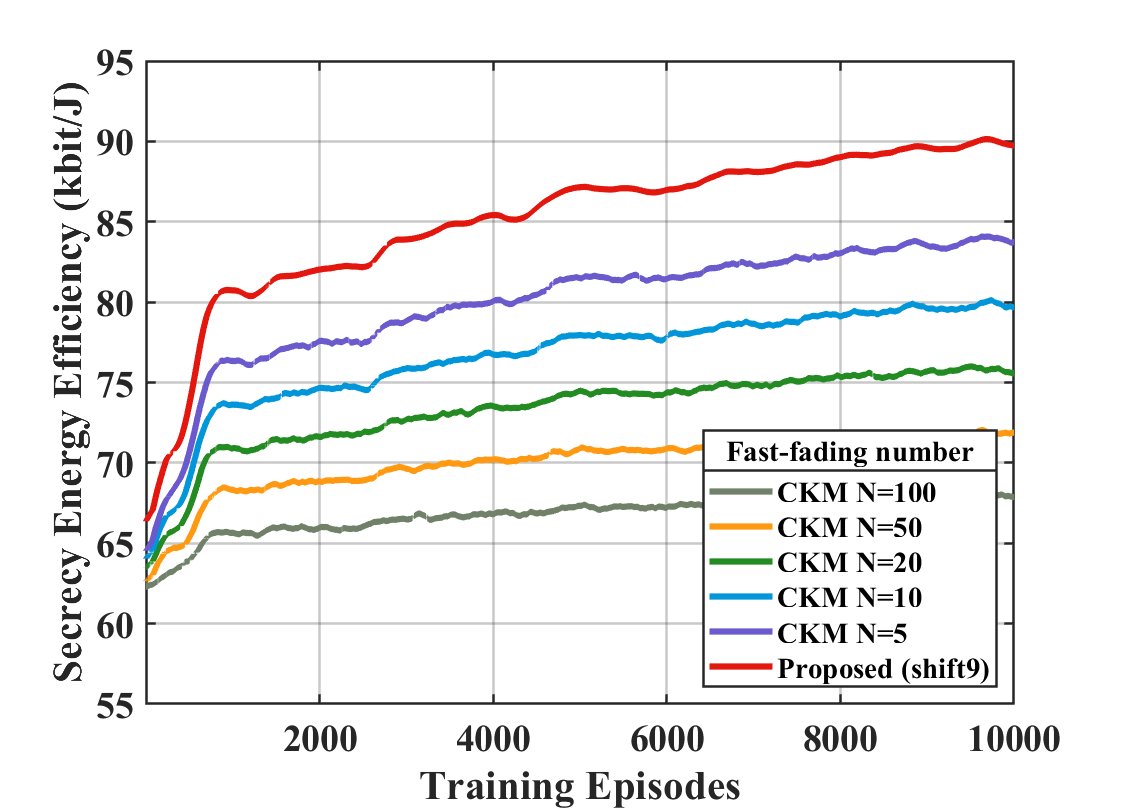}
        \caption{Impact of fast-fading realizations on algorithm performance.}
        \label{fig5b}
    \end{subfigure}\hfill
    \begin{subfigure}[t]{0.32\textwidth}
        \centering
        \includegraphics[width=\linewidth]{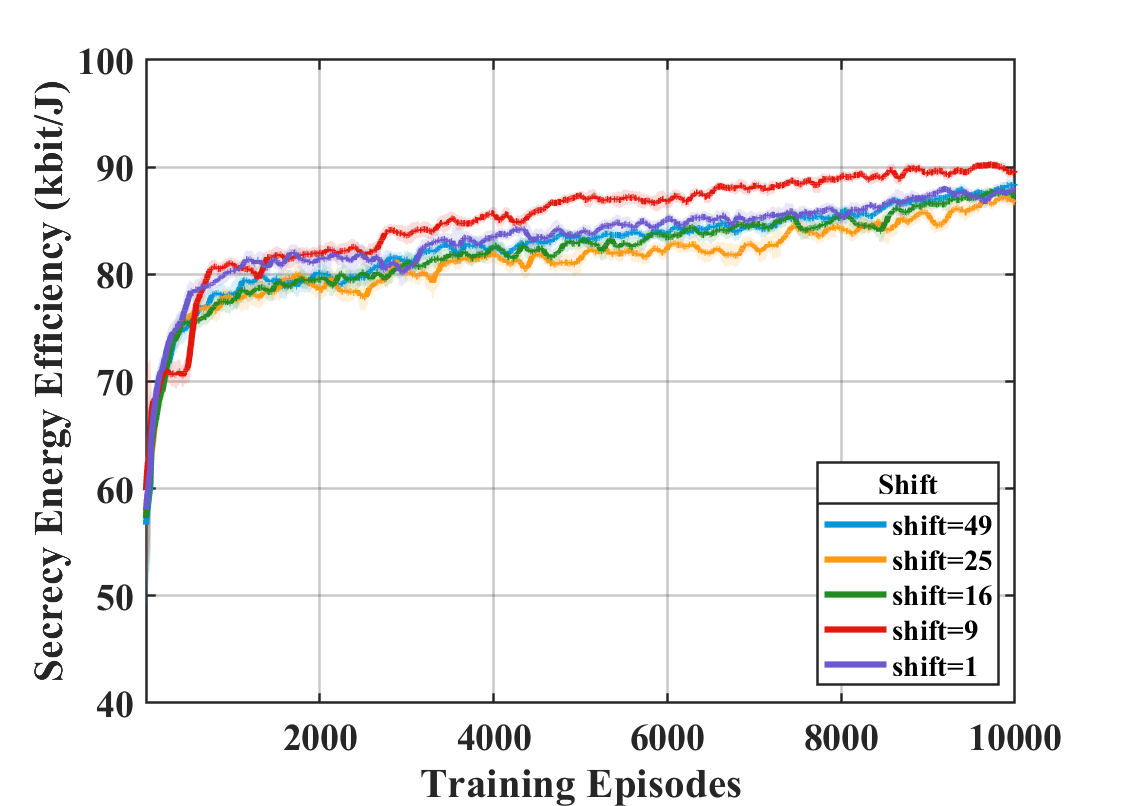}
        \caption{{Impact of the number of rigid-body shifts $N_{\mathrm{sh}}$.}}
        \label{fig5c}
    \end{subfigure}
    
    {\captionsetup{justification=raggedright,singlelinecheck=false}
        \caption{Performance comparison of the proposed algorithm under different hyperparameter settings.}
        \label{fig5}}
\end{figure*}

\begin{figure}[t]
	\centering
	\includegraphics[width=\columnwidth]{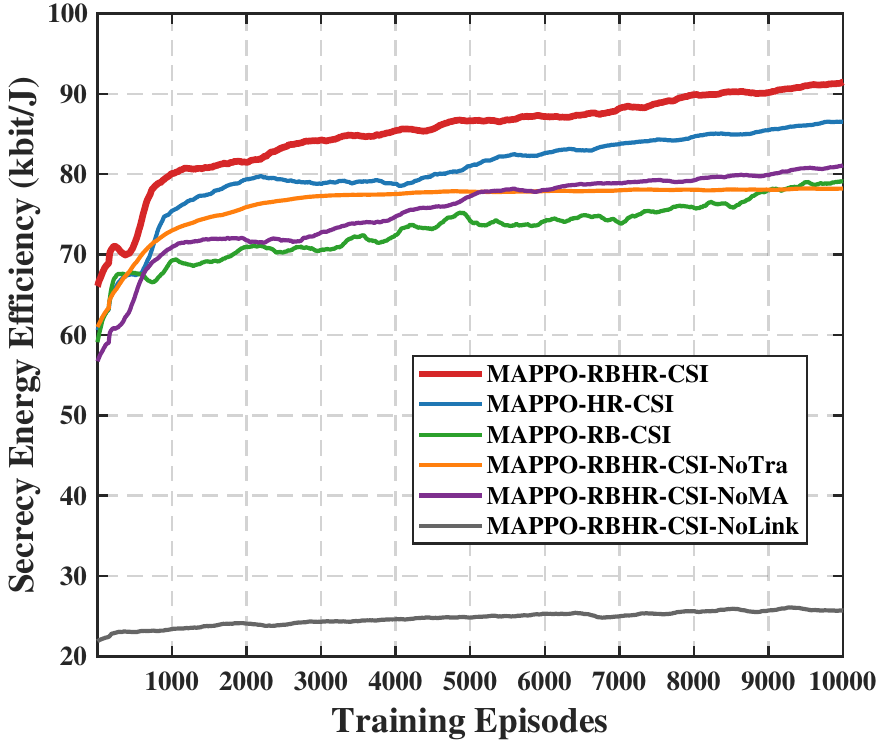}
	\caption{Comparison of SEE in Ablation Studies.}
	\label{fig6}
\end{figure}

To provide a comprehensive performance evaluation, the proposed algorithm is compared with the aforementioned benchmark schemes. SEE is adopted as the primary performance metric, while both transmit beamforming and receive combining are implemented using the minimum mean-square error (MMSE) criterion \cite{price2007communication}.

\vspace{-5mm}

\subsection{Performance Comparison}
\label{V-C}

Fig.~\ref{fig1a} illustrates the simulated scenario, where the ABS and LEO satellites extend communication coverage over the low-altitude area beyond that provided by the GBS alone. Fig.~\ref{fig1b} depicts the UAV-swarm trajectories generated by the proposed algorithm, while Fig.~\ref{fig1c} illustrates the optimized MA positions of UAV~$1$ at four representative time slots.
Fig.~\ref{fig3} illustrates the training process of the proposed algorithm. As the number of training episodes increases, both the cumulative return and SEE improve rapidly and subsequently stabilize at relatively high levels despite moderate fluctuations. Meanwhile, the critic loss remains low throughout training. Overall, these metrics exhibit favorable convergence behavior, indicating that the system SEE steadily improves and eventually stabilizes as the policy converges, thereby validating the effectiveness of the proposed algorithm.

Fig.~\ref{fig4} compares the SEE performance of the proposed algorithms and several baseline schemes during training. As the number of training episodes increases, MAPPO-RBHR-CSI rapidly improves and ultimately achieves the highest SEE. Meanwhile, MAPPO-RBHR-CKM outperforms most baseline schemes, validating the effectiveness of the proposed framework.
Under the same CSI setting, MAPPO-RBHR-CSI outperforms IPPO-CSI by approximately $4.5\%$ after convergence, indicating that cooperative multi-agent decision-making can coordinate the constituent subtasks more effectively. It also achieves an SEE gain of approximately $43\%$ over MADDPG-CSI, demonstrating the stronger policy-learning and cooperative-optimization capabilities of the adopted MAPPO mechanism for the considered hybrid discrete joint optimization problem. 
Furthermore, MAPPO-RBHR-CSI achieves an SEE improvement of approximately $12\%$ over AO-Joint-CSI. This is because alternating optimization typically optimizes one subset of variables while fixing the others, making it difficult to fully capture the long-term coupling among UAV trajectories, link selection, and MA configurations, and potentially leading to locally optimal solutions. Moreover, AO-Joint-CSI outperforms AO-FPA-CSI, AO-Straight-CSI, and AO-GN-CSI, confirming that jointly optimizing UAV trajectories, link selection, and MA positions yields substantial performance gains.

Fig.~\ref{fig5} compares the performance of the proposed algorithm across different hyperparameter settings.
{Fig.~\ref{fig5a} shows that a learning rate of $1\times10^{-4}$ attains the highest final SEE while remaining stable; this value is used in the remaining experiments.}
Fig.~\ref{fig5b} investigates the number of fast-fading realizations, where 100 realizations yield the best convergence and SEE; further increasing this number sharpens channel-statistics estimation but incurs higher computational cost, with only marginal additional gains.
{Fig.~\ref{fig5c} assesses the MA search-space size through $N_{\mathrm{sh}}$. The best result at $N_{\mathrm{sh}}=9$ indicates that too few shifts limit MA flexibility, whereas too many complicate exploration.}

\vspace{-3mm}

\subsection{Ablation Experiment}\label{V-D}

To further evaluate the proposed  algorithm, we conduct two groups of ablation studies. The first group investigates the effectiveness of the proposed rigid-body shift design and hybrid reward design. The second group examines the contributions of the three optimization variables, namely UAV trajectory, link selection, and MA position selection.
{For clarity, Shift9 denotes the proposed rigid-body MA controller with $N_{\mathrm{sh}}=9$ candidate shifts, while the suffixes NoTra, NoMA, and NoLink denote no trajectory optimization, no MA-position optimization, and no link-selection optimization, respectively.}
The resulting ablation variants are summarized as follows:

\begin{itemize}
	\item \textbf{MAPPO-HR-CSI}: This variant removes the proposed rigid-body shift design and directly optimizes the MA positions in the original high-dimensional action space.
	
	\item \textbf{MAPPO-RB-CSI}: This variant removes the hybrid individual--team reward design and adopts only individual rewards.
	
	\item \textbf{MAPPO-RBHR-CSI-NoTra}: This variant fixes the UAV trajectories as straight-line paths and optimizes only link selections and MA positions.
	
	\item \textbf{MAPPO-RBHR-CSI-NoMA}: This variant fixes the MA positions and optimizes only UAV trajectories and link selections.
	
	\item \textbf{MAPPO-RBHR-CSI-NoLink}: This variant fixes all UAVs to the GN and optimizes only UAV trajectories and MA positions.

\end{itemize}

Fig.~\ref{fig6} presents the simulation results of the ablation study. First, the effectiveness of the two proposed mechanisms is verified. As shown, removing either the Shift9 rigid-body shift mechanism  or  the individual--team hybrid reward mechanism leads to a noticeable SEE degradation. Specifically, the proposed algorithm achieves approximately $16.7\%$ and $5.8\%$ higher converged SEE than MAPPO-RB-CSI and MAPPO-HR-CSI, respectively.
This result indicates that directly embedding per-antenna MA optimization into a  hybrid action space increases the difficulty of policy learning because of the enlarged action dimension and diverse variable types, thereby validating the effectiveness of the rigid-body architecture in \eqref{eq:ma_shift_action}.
Second, scenario-level ablations are conducted. The proposed algorithm jointly optimizes the MA configuration, link selection, and UAV trajectory, whereas each of the other three variants optimizes only two of these degrees of freedom and consequently achieves lower performance. For example, disabling trajectory optimization reduces SEE by approximately $13.8\%$, while disabling link selection results in an SEE reduction of approximately $72.5\%$. These results confirm the importance of jointly optimizing all three decision variables.

\begin{table}[t]
    \caption{Comparison of the online inference complexity.}
    \label{tab:online_complexity}
    \centering
    \renewcommand{\arraystretch}{1.15}
    \begin{tabular}{lll}
        \toprule
        \textbf{Algorithm}
        & \textbf{Inference cost per slot}
        & \textbf{Mult. count} \\
        \midrule
        MAPPO-RBHR-CKM
        & $\mathcal{O}\!\left(U C_{\mathrm{actor}}^{\mathrm{shift}}\right)$
        & $\sim 7.9\times 10^{5}$ \\

        IPPO-CSI
        & $\mathcal{O}\!\left(U C_{\mathrm{actor}}^{\mathrm{full}}\right)$
        & $\sim 1.4\times 10^{6}$ \\

        MADDPG-CSI
        & $\mathcal{O}\!\left(U C_{\mathrm{actor}}^{\mathrm{full}}\right)$
        & $\sim 1.4\times 10^{6}$ \\

        AO-Joint-CSI
        & $\mathcal{O}\!\left(I_{\mathrm{ao}} U n_{\mathrm{dir}}
        n_{\mathrm{link}} M^{2} K_{\mathrm{grid}}\right)$
        & $\sim 1.9\times 10^{7}$ \\

        AO-Straight-CSI
        & $\mathcal{O}\!\left(I_{\mathrm{ao}} U n_{\mathrm{link}}
        M^{2} K_{\mathrm{grid}}\right)$
        & $\sim 2.4\times 10^{6}$ \\

        AO-GN-CSI
        & $\mathcal{O}\!\left(I_{\mathrm{ao}} U n_{\mathrm{dir}}
        M^{2} K_{\mathrm{grid}}\right)$
        & $\sim 6.3\times 10^{6}$ \\

        AO-FPA-CSI
        & $\mathcal{O}\!\left(I_{\mathrm{ao}} U n_{\mathrm{dir}}
        n_{\mathrm{link}}\right)$
        & $\sim 2.4\times 10^{4}$ \\
        \bottomrule
    \end{tabular}
\end{table}

\vspace{-3mm}

\subsection{Computational Complexity and Signal Overhead}\label{V-E}

\subsubsection{Computational Complexity}

The computational complexities of the considered algorithms are analyzed from the perspectives of offline training and online inference \cite{huang2020droo}. All schemes share the same simulation-environment kernel, whose per-slot complexity is denoted by $C_{\mathrm{env}}=\mathcal{O}\!\bigl(UM^2N+n_{\mathrm{link}}(\bar K^2N+\bar K^3)+UMK_{\mathrm{grid}}\bigr)$. This cost is primarily incurred by effective-channel computation based on singular value decomposition (SVD), regularized zero-forcing (RZF) precoding, and candidate channel-gain feature evaluation. The offline complexity analysis considers only the dominant network-update costs of the DRL methods, while the common environment-interaction cost $C_{\mathrm{env}}$ is excluded. For MAPPO-RBHR-CKM and IPPO-CSI, each training episode contains $T$ slots and each update performs $E_{\mathrm{ppo}}$ Actor--Critic optimization epochs. Accordingly, their one-time offline network-update complexities are $\mathcal{O}\!\bigl(E_{\mathrm{episode}}E_{\mathrm{ppo}}TU(C_{\mathrm{actor}}^{\mathrm{shift}}+C_{\mathrm{critic}}^{\mathrm{cen}})\bigr)$ and $\mathcal{O}\!\bigl(E_{\mathrm{episode}}E_{\mathrm{ppo}}TU(C_{\mathrm{actor}}^{\mathrm{full}}+C_{\mathrm{critic}}^{\mathrm{ind}})\bigr)$, respectively. Owing to the rigid-body shift design, the proposed method reduces the MA action-output dimension from $MK_{\mathrm{grid}}$ to $N_{\mathrm{sh}}$, thereby reducing the actor computational load. MADDPG-CSI instead performs mini-batch updates using a replay buffer, resulting in a one-time offline network-update complexity of $\mathcal{O}\!\bigl(E_{\mathrm{episode}}B_{\mathrm{batch}}U(C_{\mathrm{actor}}^{\mathrm{full}}+C_{\mathrm{critic}}^{Q})\bigr)$. These network-update costs are incurred only once offline.

During online deployment, each DRL scheme requires only one actor forward pass per UAV together with lightweight feasibility resolution; moreover, the CKM-assisted scheme does not require real-time CSI acquisition. In contrast, although AO methods require no offline training, they rely on real-time CSI and perform $I_{\mathrm{ao}}$ outer iterations at every time slot. Therefore, the proposed method trades a one-time offline training cost for substantially lower online inference complexity, making it suitable for real-time decision-making in secure UAV communications. Here, $E_{\mathrm{episode}}$, $E_{\mathrm{ppo}}$, and $I_{\mathrm{ao}}$ denote the number of training episodes, PPO epochs per update, and AO outer iterations, respectively; $U$, $T$, and $M$ denote the number of UAVs, slots per episode, and MAs per UAV, respectively; and $n_{\mathrm{dir}}$, $n_{\mathrm{link}}$, $K_{\mathrm{grid}}$, and $N_{\mathrm{sh}}$ denote the numbers of direction actions, link actions, full MA candidates, and rigid-body shift candidates, respectively. The results are summarized in Table~\ref{tab:online_complexity}.

\subsubsection{CSI Acquisition Overhead}

\begin{table}[t]
    \caption{Comparison of the CSI acquisition overhead.}
	\label{tab:csi_overhead2}
	\centering
	\renewcommand{\arraystretch}{1.15}
	\begin{tabular}{lll}
		\hline
		\textbf{Algorithm}
		& \textbf{CSI acquisition overhead}
		& \textbf{Count} \\
		\hline
		MAPPO-RBHR-CKM
		& -
		& - \\
		
		MAPPO-RBHR-CSI
		& $\mathcal{O}\!\left(E_{\mathrm{episode}} T U\right)$
		& $4.00\times10^{6}$ \\
		
		AO-Joint-CSI
		& $\mathcal{O}\!\left(R U T n_{\text{dir}}\right)$
		& $5.28\times10^{5}$ \\
		
		AO-FPA-CSI
		& $\mathcal{O}\!\left(R U T n_{\text{dir}}\right)$
		& $5.28\times10^{5}$ \\
		
		AO-GN-CSI
		& $\mathcal{O}\!\left(R U T n_{\text{dir}}\right)$
		& $5.28\times10^{5}$ \\
		
		AO-Straight-CSI
		& $\mathcal{O}\!\left(U T\right)$
		& $8.00\times10^{2}$ \\
		
		IPPO-CSI
		& $\mathcal{O}\!\left(E_{\mathrm{episode}} T U\right)$
		& $4.00\times10^{6}$ \\
		
		MADDPG-CSI
		& $\mathcal{O}\!\left(E_{\mathrm{episode}} T U\right)$
		& $4.00\times10^{6}$ \\
		\hline
	\end{tabular}
\end{table}

Table~\ref{tab:csi_overhead2} compares the total real CSI acquisition required to obtain the deployed policy, where one acquisition denotes a single agent obtaining instantaneous CSI via one interaction. 
The proposed MAPPO-RBHR-CKM replaces real channel probing with offline CKM and ephemeris data during both training and deployment, thereby requiring no real-time CSI acquisition overhead. 
By contrast, MAPPO-RBHR-CSI, IPPO-CSI, and MADDPG-CSI are all trained on perfect instantaneous CSI and must acquire CSI for all $U$ agents at every step of every training episode, giving $\mathcal{O}(E_{\mathrm{episode}}TU)$, where $E_{\mathrm{episode}}$ and $T$ denote the number of training episodes and steps per episode, respectively. The trajectory-optimizing AO methods (AO-Joint/FPA/GN-CSI) evaluate candidate positions with real CSI over $R$ restarts, $T$ slots, and $n_{\text{dir}}$ directions, yielding $\mathcal{O}(RUTn_{\text{dir}})$, whereas AO-Straight-CSI flies a fixed straight path and needs only $\mathcal{O}(UT)$. Therefore, the proposed CKM-based algorithm relies solely on the CKM and ephemeris data, fundamentally eliminating the signaling overhead of online CSI acquisition and exchange. 
Although the proposed method avoids the overhead of instantaneous CSI acquisition, the CKM requires UAV location information as its query input. This overhead is generally lower than that associated with acquiring high-dimensional instantaneous CSI  \cite{zeng2021toward}. Existing CKM studies commonly use user location as a low-dimensional, coarse-grained, and readily available query index for channel knowledge \cite{zeng2021toward}.

\vspace{-4mm}

\section{Conclusion}\label{Conclusion}


{In this paper, we investigated SAGIN-enabled secret communications for UAVs equipped with MA arrays. We maximized SEE by jointly optimizing MA translations, trajectories, and link selections. The proposed CKM-assisted MAPPO framework uses sparse offline measurements and satellite ephemerides instead of online instantaneous CSI, a low-dimensional rigid-body translation action, an individual--team reward, and a joint feasibility resolver. The resulting policy improves SEE with lower online inference and CSI-acquisition overhead than the considered baselines. Future work will study time-varying CKMs and robust optimization in dynamic-scatterer environments.}

\color{black}

\bibliographystyle{IEEEtran}
\bibliography{ref}

\begin{thebibliography}{10}
\providecommand{\url}[1]{#1}
\csname url@samestyle\endcsname
\providecommand{\newblock}{\relax}
\providecommand{\bibinfo}[2]{#2}
\providecommand{\BIBentrySTDinterwordspacing}{\spaceskip=0pt\relax}
\providecommand{\BIBentryALTinterwordstretchfactor}{4}
\providecommand{\BIBentryALTinterwordspacing}{\spaceskip=\fontdimen2\font plus
\BIBentryALTinterwordstretchfactor\fontdimen3\font minus
  \fontdimen4\font\relax}
\providecommand{\BIBforeignlanguage}[2]{{%
\expandafter\ifx\csname l@#1\endcsname\relax
\typeout{** WARNING: IEEEtran.bst: No hyphenation pattern has been}%
\typeout{** loaded for the language `#1'. Using the pattern for}%
\typeout{** the default language instead.}%
\else
\language=\csname l@#1\endcsname
\fi
#2}}
\providecommand{\BIBdecl}{\relax}
\BIBdecl
\renewcommand{\BIBentryALTinterwordstretchfactor}{4}

\bibitem{wang2024sustainable}
F.~Wang, S.~Zhang, J.~Shi, Z.~Li \emph{et~al.}, ``Sustainable {UAV} mobility
  support in integrated terrestrial and non-terrestrial networks,'' \emph{IEEE
  Trans. Wireless Commun.}, vol.~23, no.~11, pp. 17\,115--17\,128, Nov. 2024.

\bibitem{wang2025unified}
Y.~Wang, H.~Hou, X.~Yi, W.~Wang \emph{et~al.}, ``Toward unified {AI} models for
  {MU-MIMO} communications: A tensor equivariance framework,'' \emph{IEEE
  Trans. Wireless Commun.}, vol.~24, no.~12, pp. 10\,517--10\,533, 2025.

\bibitem{wan2025qos}
\BIBentryALTinterwordspacing
J.~Wan, K.~He, Y.~Wang, F.~Liu \emph{et~al.}, ``{QoS}-aware hierarchical
  reinforcement learning for joint link selection and trajectory optimization
  in {SAGIN}-supported {UAV} mobility management,'' \emph{arXiv preprint
  arXiv:2512.15119}, Dec. 2025. [Online]. Available:
  \url{https://arxiv.org/abs/2512.15119}
\BIBentrySTDinterwordspacing

\bibitem{wang2025mobile}
W.~Wang, Y.~Zhu, Y.~Wang, R.~Ding \emph{et~al.}, ``Toward mobile satellite
  internet: The fundamental limitation of wireless transmission and enabling
  technologies,'' \emph{Engineering}, vol.~54, pp. 42--51, 2025.

\bibitem{qiao2022joint}
L.~Qiao, J.~Zhang, Z.~Gao, D.~Zheng \emph{et~al.}, ``Joint activity and blind
  information detection for {UAV}-assisted massive {IoT} access,'' \emph{IEEE
  J. Sel. Areas Commun.}, vol.~40, no.~5, pp. 1489--1508, May 2022.

\bibitem{wang2026statistical}
Y.~Wang, V.~N. Ha, K.~Ntontin, H.~Yan \emph{et~al.}, ``Statistical {CSI}-based
  distributed precoding design for {OFDM}-cooperative multi-satellite
  systems,'' \emph{IEEE J. Sel. Areas Commun.}, 2026.

\bibitem{amer2020mobility}
R.~Amer, W.~Saad, and N.~Marchetti, ``Mobility in the sky: Performance and
  mobility analysis for cellular-connected uavs,'' \emph{IEEE Transactions on
  Communications}, vol.~68, no.~5, pp. 3229--3246, 2020.

\bibitem{zhu2023modeling}
L.~Zhu, W.~Ma, and R.~Zhang, ``Modeling and performance analysis for movable
  antenna enabled wireless communications,'' \emph{IEEE Trans. Wireless
  Commun.}, vol.~23, no.~6, pp. 6234--6250, Jun. 2024.

\bibitem{cao2025channel}
S.~Cao, L.~Zhu, Z.~Xiao, and B.~Ning, ``Channel estimation for movable antenna
  aided wideband communication systems,'' in \emph{Proc. IEEE Wireless Commun.
  Netw. Conf. (WCNC)}, Milan, Italy, Mar. 2025, pp. 1--6.

\bibitem{zhu2024historical}
\BIBentryALTinterwordspacing
L.~Zhu and K.-K. Wong, ``Historical review of fluid antenna and movable
  antenna,'' \emph{arXiv preprint arXiv:2401.02362}, Jan. 2024. [Online].
  Available: \url{https://arxiv.org/abs/2401.02362}
\BIBentrySTDinterwordspacing

\bibitem{zheng2024flexible}
J.~Zheng, J.~Zhang, H.~Du, D.~Niyato \emph{et~al.}, ``Flexible-position {MIMO}
  for wireless communications: Fundamentals, challenges, and future
  directions,'' \emph{IEEE Wireless Commun.}, vol.~31, no.~5, pp. 18--26, Oct.
  2024.

\bibitem{shao2025network}
X.~Shao, R.~Zhang, Q.~Jiang, and R.~Schober, ``{6D} movable antenna enhanced
  wireless network via discrete position and rotation optimization,''
  \emph{IEEE J. Sel. Areas Commun.}, vol.~43, no.~3, pp. 674--687, Mar. 2025.

\bibitem{lei2025multi}
H.~Lei, D.~Meng, H.~Ran, K.-H. Park \emph{et~al.}, ``Multi-uav trajectory
  design for fair and secure communication,'' \emph{IEEE Transactions on
  Cognitive Communications and Networking}, vol.~11, no.~3, pp. 1966--1980,
  2025.

\bibitem{jiang2021covert}
X.~Jiang, X.~Chen, J.~Tang, N.~Zhao \emph{et~al.}, ``Covert communication in
  {UAV}-assisted air-ground networks,'' \emph{IEEE Wireless Commun.}, vol.~28,
  no.~4, pp. 190--197, Aug. 2021.

\bibitem{zhang2024toward}
T.~Zhang, Y.~Xu, J.~Zhao, J.~Xue \emph{et~al.}, ``Toward handover-free mobility
  management in {FD}-{RAN}: Architecture, challenges, and solutions,''
  \emph{IEEE Netw.}, vol.~38, no.~6, pp. 433--442, Nov. 2024.

\bibitem{yang2023dqn}
J.~Yang, Z.~Xiao, H.~Cui, J.~Zhao \emph{et~al.}, ``{DQN}-{ALrM}-based
  intelligent handover method for satellite-ground integrated network,''
  \emph{IEEE Trans. Cogn. Commun. Netw.}, vol.~9, no.~4, pp. 977--990, Aug.
  2023.

\bibitem{xu2017modeling}
X.~Xu, Z.~Sun, X.~Dai, T.~Svensson \emph{et~al.}, ``Modeling and analyzing the
  cross-tier handover in heterogeneous networks,'' \emph{IEEE Trans. Wireless
  Commun.}, vol.~16, no.~12, pp. 7859--7869, Dec. 2017.

\bibitem{liu2023user}
Q.~Liu, X.~Li, H.~Ji, and H.~Zhang, ``User grouping-based beam handover scheme
  with load-balancing for {LEO} satellite networks,'' in \emph{Proc. IEEE Glob.
  Commun. Conf. (GLOBECOM)}, Kuala Lumpur, Malaysia, Dec. 2023, pp. 3965--3970.

\bibitem{zhang2018cellular}
S.~Zhang, Y.~Zeng, and R.~Zhang, ``Cellular-enabled {UAV} communication: A
  connectivity-constrained trajectory optimization perspective,'' \emph{IEEE
  Trans. Commun.}, vol.~67, no.~3, pp. 2580--2604, Mar. 2019.

\bibitem{bulut2018trajectory}
E.~Bulut and I.~Guevenc, ``Trajectory optimization for cellular-connected
  {UAVs} with disconnectivity constraint,'' in \emph{Proc. IEEE Int. Conf.
  Commun. Workshops (ICC Workshops)}, Kansas City, MO, USA, May 2018, pp. 1--6.

\bibitem{zhang2019trajectory}
S.~Zhang and R.~Zhang, ``Trajectory design for cellular-connected {UAV} under
  outage duration constraint,'' in \emph{Proc. IEEE Int. Conf. Commun. (ICC)},
  Shanghai, China, May 2019, pp. 1--6.

\bibitem{zeng2021simultaneous}
Y.~Zeng, X.~Xu, S.~Jin, and R.~Zhang, ``Simultaneous navigation and radio
  mapping for cellular-connected {UAV} with deep reinforcement learning,''
  \emph{IEEE Trans. Wireless Commun.}, vol.~20, no.~7, pp. 4205--4220, Jul.
  2021.

\bibitem{zhan2022energy}
C.~Zhan and Y.~Zeng, ``Energy minimization for cellular-connected {UAV}: From
  optimization to deep reinforcement learning,'' \emph{IEEE Trans. Wireless
  Commun.}, vol.~21, no.~7, pp. 5541--5555, Jul. 2022.

\bibitem{zhu2024movable}
L.~Zhu, W.~Ma, and R.~Zhang, ``Movable antennas for wireless communication:
  Opportunities and challenges,'' \emph{IEEE Commun. Mag.}, vol.~62, no.~6, pp.
  114--120, Jun. 2024.

\bibitem{zhu2023movable}
------, ``Movable-antenna array enhanced beamforming: Achieving full array gain
  with null steering,'' \emph{IEEE Commun. Lett.}, vol.~27, no.~12, pp.
  3340--3344, Dec. 2023.

\bibitem{bai2025movable}
Y.~Bai, B.~Xie, R.~Zhu, Z.~Chang \emph{et~al.}, ``Movable antenna-equipped
  {UAV} for data collection in backscatter sensor networks: A deep
  reinforcement learning approach,'' in \emph{Proc. IEEE Int. Conf. Commun.
  (ICC)}, Montreal, QC, Canada, Jun. 2025, pp. 6560--6565.

\bibitem{liu2025uav_dup1}
W.~Liu, X.~Zhang, H.~Xing, J.~Ren \emph{et~al.}, ``{UAV}-enabled wireless
  networks with movable-antenna array: Flexible beamforming and trajectory
  design,'' \emph{IEEE Wireless Commun. Lett.}, vol.~14, no.~3, pp. 566--570,
  Mar. 2025.

\bibitem{wen2026flexible}
Z.~Wen, M.~Liu, W.~X. Zheng, and J.~Zhang, ``Flexible position antenna enabled
  resource optimization for {UAV} covert communication,'' \emph{IEEE
  Transactions on Network Science and Engineering}, vol.~13, pp. 4455--4471,
  2026.

\bibitem{kim2026energy}
S.~Kim, J.~Gong, and J.~Kang, ``Energy-efficient secure communications via
  joint optimization of {UAV} trajectory and movable-antenna array
  beamforming,'' \emph{IEEE Wireless Commun. Lett.}, vol.~15, pp. 1210--1214,
  2026.

\bibitem{wan2026channel}
J.~Wan, H.~Hou, J.~Zhuang, W.~Wang \emph{et~al.}, ``A channel knowledge
  map-driven two-stage coordinated user scheduling in multi-cell massive {MIMO}
  systems,'' \emph{IEEE Trans. Commun.}, 2026.

\bibitem{zeng2021toward}
Y.~Zeng and X.~Xu, ``Toward environment-aware {6G} communications via channel
  knowledge map,'' \emph{IEEE Wireless Commun.}, vol.~28, no.~3, pp. 84--91,
  Jun. 2021.

\bibitem{hu2020cooperative}
J.~Hu, H.~Zhang, L.~Song, R.~Schober \emph{et~al.}, ``Cooperative internet of
  {UAVs}: Distributed trajectory design by multi-agent deep reinforcement
  learning,'' \emph{IEEE Transactions on Communications}, vol.~68, no.~11, pp.
  6807--6821, 2020.

\bibitem{wu2020cellular}
F.~Wu, H.~Zhang, J.~Wu, and L.~Song, ``Cellular {UAV}-to-device communications:
  Trajectory design and mode selection by multi-agent deep reinforcement
  learning,'' \emph{IEEE Transactions on Communications}, vol.~68, no.~7, pp.
  4175--4189, 2020.

\bibitem{kim2022non}
S.~Kim, M.~Kim, J.~Y. Ryu, J.~Lee \emph{et~al.}, ``Non-terrestrial networks for
  {UAVs}: Base station service provisioning schemes with antenna tilt,''
  \emph{IEEE Access}, vol.~10, pp. 41\,537--41\,550, 2022.

\bibitem{roste2026terminals}
T.~R{\o}ste, O.~Gutteberg, and H.~C. Haugli, ``User terminals with multipaneled
  phased arrays and omnidirectional antennas applied in an elliptical medium
  earth orbit satellite communication system for the arctic,'' \emph{Frontiers
  in Antennas and Propagation}, vol.~4, p. 1764792, 2026.

\bibitem{khuwaja2018survey}
A.~A. Khuwaja, Y.~Chen, N.~Zhao, M.-S. Alouini \emph{et~al.}, ``A survey of
  channel modeling for {UAV} communications,'' \emph{IEEE Commun. Surv.
  Tutor.}, vol.~20, no.~4, pp. 2804--2821, 2018.

\bibitem{ma2024mimo}
W.~Ma, L.~Zhu, and R.~Zhang, ``{MIMO} capacity characterization for movable
  antenna systems,'' \emph{IEEE Trans. Wireless Commun.}, vol.~23, no.~4, pp.
  3392--3407, Apr. 2024.

\bibitem{xiao2022antenna}
Z.~Xiao, Z.~Han, A.~Nallanathan, O.~A. Dobre \emph{et~al.}, ``Antenna array
  enabled space/air/ground communications and networking for {6G},'' \emph{IEEE
  J. Sel. Areas Commun.}, vol.~40, no.~10, pp. 2773--2804, Oct. 2022.

\bibitem{zhang2019securing}
G.~Zhang, Q.~Wu, M.~Cui, and R.~Zhang, ``Securing {UAV} communications via
  joint trajectory and power control,'' \emph{IEEE Trans. Wireless Commun.},
  vol.~18, no.~2, pp. 1376--1389, Feb. 2019.

\bibitem{zeng2019energy}
Y.~Zeng, J.~Xu, and R.~Zhang, ``Energy minimization for wireless communication
  with rotary-wing {UAV},'' \emph{IEEE Trans. Wireless Commun.}, vol.~18,
  no.~4, pp. 2329--2345, Apr. 2019.

\bibitem{wu2023environment}
D.~Wu, Y.~Zeng, S.~Jin, and R.~Zhang, ``Environment-aware hybrid beamforming by
  leveraging channel knowledge map,'' \emph{IEEE Trans. Wireless Commun.},
  vol.~23, no.~5, pp. 4990--5005, May 2024.

\bibitem{cheng2026channel}
N.~Cheng, S.~Yang, R.~Sun, Z.~Yin \emph{et~al.}, ``Channel knowledge
  map-enabled {6D} movable antenna systems with kinematic constraints: A
  manifold optimization approach,'' \emph{IEEE Trans. Wireless Commun.},
  vol.~25, pp. 8968--8981, 2026.

\bibitem{li2023channel}
H.~Li, P.~Li, G.~Cheng, J.~Xu \emph{et~al.}, ``Channel knowledge map
  ({CKM})-assisted multi-{UAV} wireless network: {CKM} construction and {UAV}
  placement,'' \emph{J. Commun. Inf. Netw.}, vol.~8, no.~3, pp. 256--270, Sept.
  2023.

\bibitem{Mei2024Graph}
W.~Mei, X.~Wei, B.~Ning, Z.~Chen \emph{et~al.}, ``Movable-antenna position
  optimization: A graph-based approach,'' \emph{IEEE Wireless Commun. Lett.},
  vol.~13, no.~7, pp. 1853--1857, Jul. 2024.

\bibitem{huang2022invalid}
S.~Huang and S.~Onta{\~n}{\'o}n, ``A closer look at invalid action masking in
  policy gradient algorithms,'' in \emph{Proc. Int. Florida Artif. Intell. Res.
  Soc. Conf. (FLAIRS)}, vol.~35, 2022.

\bibitem{wang2025multi}
Y.~Wang, Y.~Zhu, V.~N. Ha, W.~Wang \emph{et~al.}, ``Multi-satellite
  multi-stream beamspace massive {MIMO} transmission,'' \emph{arXiv preprint
  arXiv:2512.21998}, 2025.

\bibitem{deWitt2020IPPO}
\BIBentryALTinterwordspacing
C.~S. de~Witt, T.~Gupta, D.~Makoviichuk, V.~Makoviychuk \emph{et~al.}, ``Is
  independent learning all you need in the {StarCraft} multi-agent challenge?''
  \emph{arXiv preprint arXiv:2011.09533}, Nov. 2020. [Online]. Available:
  \url{https://arxiv.org/abs/2011.09533}
\BIBentrySTDinterwordspacing

\bibitem{lowe2017maddpg}
R.~Lowe, Y.~Wu, A.~Tamar, J.~Harb \emph{et~al.}, ``Multi-agent actor-critic for
  mixed cooperative-competitive environments,'' in \emph{Proc. Adv. Neural Inf.
  Process. Syst. (NeurIPS)}, 2017, pp. 6379--6390.

\bibitem{price2007communication}
R.~Price and P.~E. Green, ``A communication technique for multipath channels,''
  \emph{Proc. {IRE}}, vol.~46, no.~3, pp. 555--570, Mar. 1958.

\bibitem{huang2020droo}
L.~Huang, S.~Bi, and Y.-J.~A. Zhang, ``Deep reinforcement learning for online
  computation offloading in wireless powered mobile-edge computing networks,''
  \emph{IEEE Trans. Mobile Comput.}, vol.~19, no.~11, pp. 2581--2593, Nov.
  2020.

\end{thebibliography}

\end{document}